\pdfoutput=1

\documentclass[12pt,english,a4paper]{article}

\usepackage{amsmath,amsfonts,amssymb,babel,slashed,empheq,tensor}
\usepackage[utf8]{inputenc}
\usepackage{cite}
\usepackage[pdftex,breaklinks]{hyperref}
\usepackage{amsmath}
\usepackage{amsfonts}
\usepackage{array}
\usepackage{amssymb}
\usepackage{latexsym}
\usepackage{mathrsfs}
\usepackage{braket}		
\usepackage{setspace}
\usepackage{mathrsfs}
\usepackage{graphicx}
\usepackage{xcolor}
\usepackage{mathtools}
\usepackage{slashed}
\usepackage{overpic}
\usepackage[all]{xy}
\usepackage{tikz-cd}
\usepackage{slashed}
\usepackage{empheq}
\usepackage{float}
\usepackage{mathtools}
\usepackage{tikz}
\usepackage{bm}
\usepackage{stackengine}
\usepackage{scalerel}
\usepackage{pgfplots}
\pgfplotsset{compat=1.18}
\def\rlwd{.9pt}
\def\lhexbrace{\kern1pt%
\setstackgap{S}{0pt}\def\stackalignment{l}
\ThisStyle{\scalerel*{%
  \stackunder[-\rlwd]{%
    \stackon[-\rlwd]{\roundrule{\rlwd}{4pt}}{\rotatebox{60}{\roundrule{4pt}{\rlwd}}}%
  }{\rotatebox{-60}{\roundrule{4pt}{\rlwd}}}%
}{\SavedStyle[}}}
\def\rhexbrace{%
\setstackgap{S}{0pt}\def\stackalignment{r}
\ThisStyle{\scalerel*{%
  \stackunder[-\rlwd]{%
    \stackon[-\rlwd]{\roundrule{\rlwd}{4pt}}{\rotatebox{-60}{\roundrule{4pt}{\rlwd}}}%
  }{\rotatebox{60}{\roundrule{4pt}{\rlwd}}}%
}{\SavedStyle[}}\kern1pt}

\makeatletter

\def\bra#1{\left\langle#1\right|}
\def\ket#1{\left|#1\right\rangle}

\newcommand{\B}{\mathbb{B}}

\newcommand{\mso}{\mathfrak{so}}

\newcommand{\Tr}{\text{Tr}}

\newcommand{\tn}{\mathsf{n}}

\newcommand{\nn}{\nonumber}

 \def\one{\mbox{1 \kern-.59em {\rm l}}}

\newcommand{\cK}
{\mathcal{K}}
\newcommand{\cS}
{\mathcal{S}}

 \def\cQ{{\cal Q}}
 \def\cM{{\cal M}}
 \def\cH{{\cal H}}
 \def\cF{{\cal F}}
 \def\cC{{\cal C}}
 \def\cO{{\cal O}}

\newcommand{\End}{\mathrm{End}}
\newcommand{\tr}{\mathrm{tr}}

\def\R{{\mathbb R}} \def\C{{\mathbb C}} \def\N{{\mathbb N}}

\newcommand{\hs}{\mathfrak{hs}}

\newcommand{\msu}{\mathfrak{su}}

\newcommand{\und}{\underline}
\newcommand{\oveq}[2]{\overset{#1}{\overbrace{#2}}}

\newcommand{\nket}[1]{{\bm{\Vert} #1 \bm{\rangle}}}

\newcommand{\nbraket}[2]{\bm{\langle} {#1} \bm{\Vert\!\!\Vert}{#2}\bm{\rangle}}

\def\a{\alpha}  \def\b{\beta}

\def\t{\tau} 
\def\L{\Lambda}

 \allowdisplaybreaks[3]

\makeatother

\begin{document}

\renewcommand{\title}[1]{\vspace{10mm}\noindent{\Large{\bf#1}}\vspace{8mm}}
\newcommand{\authors}[1]{\noindent{\large #1}\vspace{5mm}}
\newcommand{\address}[1]{{\itshape #1\vspace{2mm}}}


\begin{titlepage}
\begin{flushright}
 UWThPh 2025-4
\end{flushright}
\begin{center}

\title{ {\Large Minimal covariant quantum space-time }  }

\vskip 3mm

\authors{Alessandro Manta and Harold C.\ Steinacker}

\makeatletter{\renewcommand*{\@makefnmark}{}
\footnotetext{E-mail: \texttt{alessandro.manta@univie.ac.at}, \texttt{harold.steinacker@univie.ac.at}}\makeatother}

\vskip 3mm

 \address{
{\it Faculty of Physics, University of Vienna\\
Boltzmanngasse 5, 1090 Vienna, Austria  } }

\bigskip

\vskip 1.4cm

\textbf{Abstract}
\vskip 3mm

\begin{minipage}{14.8cm}%
\vskip 3mm

We discuss minimal covariant quantum space-time $\cM^{1,3}_0$, which is
defined through the minimal doubleton representation of $\mso(4,2)$.
 An elementary definition in terms of generators and relations is  given.
This space is shown to admit a semi-classical interpretation as
 quantized twistor space $\C P^{1,2}$, viewed as
a quantized $S^2$-bundle over a 3+1-dimensional $k=-1$ FLRW space-time. In particular we find an over-complete set of
 (quasi-) coherent states, with a large hierarchy between the uncertainty scale and the geometric curvature scale.
 This provides an interesting background for the IKKT model, leading to a $\hs$-extended gravitational gauge theory, which is free of ghosts due to the constraints on phase space arising from the doubleton representation.

\end{minipage}

\end{center}

\end{titlepage}

\tableofcontents



\section{Introduction}

It is generally expected that the classical concept of space-time 
ceases to make sense below the Planck scale, where quantum effects to gravity become important.
The mathematical description of the underlying "quantum geometry" is however not clear. One current idea is that holography may provide a description of quantum geometry, cf. \cite{Maldacena:1997re,Hartnoll:2024csr,Komatsu:2024bop}. This geometry would then be described through the deep quantum regime of a strongly coupled gauge theory. 

A different possible description of quantum space-time is through some  noncommutative algebraic structure.
This approach has the great advantage that it can be realized within the weakly-coupled regime of some underlying (generalized) gauge theory, such as a matrix model. That idea has been discussed extensively in the literature, cf. \cite{Doplicher:1994tu,Grosse:1995ar,Szabo:2001kg,Douglas:2001ba}; the main difficulty lies in finding a specific realization of such a quantum space-time which is physically acceptable, and leads to (near-) realistic physics in suitable models.

A particularly promising family of such quantum space-times was put forward in \cite{Steinacker:2017bhb,Sperling:2019xar}, denoted as covariant quantum space-time\footnote{For related proposals for covariant quantum space-times see e.g. \cite{Snyder:1946qz, Yang:1947ud, Heckman:2014xha, Buric:2017yes, Gazeau:2009mi}.} $\cM^{1,3}_n$.
They are based on the minimal series of unitary 
representations $\cH_n$ of $SO(4,2)$ known as "doubletons" or "minireps"  \cite{Fernando:2009fq}, for $n\in\N$.
It was shown in \cite{Sperling:2019xar} that  at least for large $n\gg 1$,
these spaces admit a semi-classical regime as  maximally homogeneous and isotropic space-times,
more precisely as quantized $S^2$ bundles over space-time. Considered as a background  in  Yang-Mills matrix models, the space  
acquires an $SO(3,1)$-invariant $k=-1$ FLRW metric with Big Bounce, and leads to an
 interesting higher-spin gauge theory. For this theory to be almost-local and  to avoid strong UV/IR mixing, it is essential to choose the 
 maximally supersymmetric IKKT matrix model \cite{Ishibashi:1996xs}. This is related to on-going efforts to understand the emergence of 3+1-dimensional space-time
 as dynamical vacua of this and related models \cite{Nishimura:2019qal,Hirasawa:2024dht,Asano:2024def,Chou:2025moy,Brahma:2022dsd}.

In the present paper, we establish that already the minimal case $n=0$ in this family of covariant quantum spaces admits a satisfactory semi-classical regime, clarifying certain open issues in previous works.
This is non-trivial, because the standard group-theoretical ("Perelomov") coherent states based on the lowest weight states (as considered in previous works) are {\bf not} useful in the present context: they saturate the  uncertainty relations on $\cM^{1,3}_n$ in a very unbalanced way, and resolve the internal $S^2$ fiber only for $n \gg 1$\footnote{For a different treatment of quasi-coherent states on such spaces see \cite{Hasebe_2021}.}.

We clarify this issue by explicitly constructing a set  of (quasi-) coherent states for minimal $\cM^{1,3}_0$ with good localization properties, which saturate the geometrical uncertainty relations, and resolve the  $S^2$ fiber over space-time.
This is achieved in two  ways: 
\begin{enumerate}
\item
An analytical construction of such states is given based on the oscillator construction of $\cH_0$. This construction also justifies the interpretation in terms of quantized twistor space $\C P^{1,2}$.

\item
 Quasi-coherent states are obtained numerically, using the framework of quasi-coherent states established in \cite{Steinacker:2020nva}.
 
\end{enumerate}
 We verify numerically that these two constructions essentially agree with each other. In particular, they resolve the geometry up to a UV scale $L_{NC}$, as expected from the underlying commutation relations.
 A large separation between the geometrical IR scale and this UV scale arises at late times, which makes $\cM_0^{1,3}$ an interesting candidate for a realistic quantum space-time, with finitely many dof per volume and a large covariance group.
 These states are over-complete and manifestly covariant under space-like $SO(4)$ transformations, which act transitively on the space-like slices of the $S^2$ bundle over space-time.
On the other hand, the states are not manifestly covariant under boosts. This should be expected, since boosts are not isometries in a generic FLRW geometry\footnote{Lorentz boosts are expected to
be recovered via volume-preserving diffeomorphisms, which are realized by $U(\cH_0)$ transformations.}.

These findings imply that previous results on the geometry and physics arising on the covariant quantum spaces with large $n$  \cite{Sperling:2019xar,Steinacker:2019awe,Steinacker:2021yxt}  ff.
 apply also to the minimal $n=0$ case.
 In particular, a tower of higher-spin modes\footnote{This corrects some  statements on the $\hs$ cutoff in previous works.} arises even for $n=0$, and it has a "local" cutoff evolving under the cosmic evolution.
Furthermore,
 we also give an elementary definition of  minimal $\cM^{1,3}_0$ via generators and relations, which can be interpreted in terms of quantized phase space with constraints. This makes that quantum space more accessible without  advanced group theory, and should facilitate future work.

\section{Algebraic structure of minimal covariant quantum space-time}

\subsection{Minimal doubleton representation $\cH_0$ of $SO(4,2)$}
\label{sec:min-H0}

Let $\eta^{ab} = {\rm diag}(-1,1,1,1,1,-1)$ be the invariant metric of $SO(4,2)$, and
$\cM^{ab}$ be hermitian generators of $SO(4,2)$ or its universal cover $SU(2,2)$,  which satisfy
\begin{align}
  [\cM_{ab},\cM_{cd}] &=i(\eta_{ac}\cM_{bd} - \eta_{ad}\cM_{bc} - \eta_{bc}\cM_{ad} + \eta_{bd}\cM_{ac}) \ .
 \label{M-M-relations-noncompact}
\end{align}
We choose the minimal (discrete series) positive-energy unitary irrep $\cH_0$ of $SO(4,2)$ \cite{Fernando:2009fq}, as detailed in section \ref{sec:osc-CP12}.
We then define the hermitian generators
\begin{subequations}\label{eq:hermitian}
    \begin{align}
 X^a &:= r\cM^{a 5}, \qquad a = 0,...,4  \nn\\
 T^\mu &:= \frac 1r  \cM^{\mu 4}, \qquad \mu = 0,...,3
\end{align} 
\end{subequations}
which transform as vectors under $SO(4,1)$ and $SO(3,1)$, respectively. Moreover,
\begin{align}
     [X^a,X^b] &= -i r^2\cM^{ab}  =: i\Theta^{ab} \ .
\end{align}
The spectrum of $X^0$ 
is then positive and discrete 
\begin{align}
\label{spec-X0}
 {\rm spec}(X^0) =r\,{\rm spec }(\cM^{05}) = r\{1, 2, 3, ... \}
\end{align}
where the 
eigenspace with lowest eigenvalue of $X^{0}$ is one-dimensional. These generators satisfy the following relations  (cf. \cite{Sperling:2019xar})
\begin{subequations}
\label{constraints-CP12}
    \begin{align}
    X_a X^a &\equiv -X_0^2 + \sum_{i=1}^4 X_i^2 =  r^2 \one\,,   \label{radius-constraint}\\
    T_\mu T^\mu &= -\frac 1{r^2} \one + \frac{X_4^2}{r^4} \,,\label{Tsquare-id}\\
    X_\mu T^\mu + T_\mu X^\mu &= 0 \label{TX-id}
\end{align}
\end{subequations}
as well as
\begin{subequations}
    \begin{align}
 \eta_{cc'}\cM^{ac} \cM^{bc'} + (a \leftrightarrow b)
  &= - 2\!\eta^{ab} + r^{-2}\left(X^a X^b + X^b X^a \right)  \
 \label{Theta-constraint}  \\
    \cM^{ab} X_b + X_b\cM^{ab}  &= 0  \label{MX-id} \\
    \varepsilon_{abcdef} \cM^{ab}\cM^{cd} &= 0 \,, \label{MM-eps-so42-id} \\
    \varepsilon_{abcde} \cM^{ab}\cM^{cd} &= 0  \,,\label{selfdual-constraint-1}\\
    \varepsilon_{abcde} \cM^{ab}X^{c} &= 0  \,.
    \label{selfdual-constraint-2}
\end{align}
\end{subequations}
Here latin indices run from 0 to 4, except in \eqref{MM-eps-so42-id} where they run from 0 to 5.
Greek indices always run from $0$ to $3$.
The radial constraint \eqref{radius-constraint}  suggests an interpretation in terms of a 4-dimensional Euclidean hyperboloid $H^4 \subset \R^{1,4}$.
We observe that \eqref{radius-constraint} and \eqref{Tsquare-id}  imply 
\begin{align}
    r^2  T_\mu T^\mu &=  - r^{-2}  X_\mu X^\mu \ .
    \label{TT-XX-relation}
\end{align}
We will also need the following commutation relations
\begin{subequations}
    \begin{align}
    [T^\mu,X^\nu] &= \frac{i}{r} X^4\eta^{\mu\nu}\,,   \\
    [X^4,X^\mu] 
    &= i r^3 T^\mu, \qquad
     [X^4,T^\mu] = \frac{i}{r} X^\mu\, \\
     [T^\mu ,T^\nu] &= \frac{i}{r^2} \cM^{\mu\nu}.
\end{align}
\label{T-X-CR}
\end{subequations}
Some comments on the interpretation are in order.
It is well-known that  $SO(4,2)$
can be interpreted as conformal group acting on Minkowski space. From that point of view, 
the Hilbert space $\cH_0$ 
can be interpreted as space of massless scalar fields 
on Minkowski space \cite{Mack:1975je,Fernando:2009fq}.
Such a field on Minkowski space is uniquely specified by its initial conditions on a 3-dimensional Cauchy surface. 
This suggests that $\End(\cH_0) \cong \cH_0 \otimes \cH_0^*$ can be interpreted in terms of (quantized) functions on a 6-dimensional space.

In the present context, we will interpret $SU(2,2) \cong SO(4,2)$ as symmetry group acting on twistor space $\C P^{1,2}$, which is viewed as $S^2$ bundle over $H^4$. Accordingly, $\End(\cH_0)$ will be interpreted in terms of (quantized) functions on 6-dimensional twistor space.
Some relevant facts about the structure of the Lie algebra $\mso(4,2)$ are collected in appendix \ref{sec:mso42-structure}.

The structure of the relations 
\eqref{T-X-CR}
is reminiscent of Snyder-type noncommutative (phase) spaces \cite{Snyder:1946qz,Yang:1947ud,Meljanac:2022qhp}, cf. also \cite{Heckman:2014xha}. A crucial distinction though is the specific choice of representation, which leads to further constraints as elaborated below. Moreover,
the $T^\mu$ generators will {\em not} be interpreted as momenta but as generators of an internal sphere $S^2$; only the $[T^\mu,\ \cdot \ ]$ should be interpreted as momenta. This is essential to obtain a consistent field theory which is free of ghosts \cite{Steinacker:2019awe}.

\subsection{Harmonic mode analysis}

We are primarily interested in the space of observables i.e. hermitian operators in $\End(\cH_0)$ and their relation to classical functions on an underlying space.
One way to identify that  space is by classifying the  independent (symmetric) polynomials.
This can be done either from the $SO(4,1)$  point of view related to   $H^4$, or from the $SO(3,1)$ point of view  adapted to the space-time $\cM^{1,3}$. 
Most of the following can be easily generalized to the case $n>0$.

\paragraph{$SO(4,1)$ point of view.}

Like all doubleton irreps of $SO(4,2)$, the minimal doubleton $\cH_0$  remains irreducible as a representation of $SO(4,1)$.
The quadratic $SO(4,1)$ Casimir is obtained from \eqref{Theta-constraint}
\begin{align}
   \frac 12 \cM^{ab} \cM_{ab}
  &=  - \frac 52 + \frac 1{2r^2} X^a X_a = -2
\end{align}
and the quartic Casimir is 
$C_4 =  (\varepsilon \cM \cM)^2 = 0$.
This means that $\cH_0$ corresponds to the minimal discrete irrep $\Pi_{1,0}$, of $SO(4,1)$ in the notation of \cite{dixmier1961representations}.

\paragraph{$SO(3,1)$ point  of view.}

The quadratic $SO(3,1)$ Casimir can be written using \eqref{Theta-constraint} 
\begin{align}
\cM^{\mu\nu} \cM_{\mu\nu} + 2  \cM^{\mu 4} \cM_{\mu 4}
  = - 5 + r^{-2} X^a X_a = -4
\end{align}
as 
\begin{align}
\label{SO31-casimir-1}
   \frac 12\cM^{\mu\nu} \cM_{\mu\nu} 
  = - 1 - \frac 1{r^2} X_4^2 \leq -1 \ .
\end{align}
The other Casimir 
\begin{align}
\varepsilon_{\mu\nu\rho\sigma} \cM^{\mu\nu} \cM^{\rho\sigma} = 0 
   \label{SO31-casimir-2}
\end{align}
vanishes due to \eqref{selfdual-constraint-1}. 

We can now discuss the space of vector or tensor operators up to quadratic order:

\begin{itemize}
 \item {\bf vector operators under $SO(3,1)$, $SO(4)$ and $SO(3)$}

There are clearly two independent vector operators of $SO(3,1)$, given by $X^\mu$ and $T^\mu$, as in the $n>0$ case. 
The other candidates vanish because
\begin{align}
   0 &=  \varepsilon_{\mu\nu\rho\sigma} \cM^{\mu\nu}X^\sigma  \propto  \varepsilon_{\mu\nu\rho\sigma} X^\mu X^\nu X^\sigma\nn\\
    0 &=  \varepsilon_{\mu\nu\rho\sigma} \cM^{\mu\nu}T^\sigma 
 \propto  \varepsilon_{\mu\nu\rho\sigma} T^\mu T^\nu T^\sigma
 \label{eps-M-X-T}
\end{align}
due to \eqref{selfdual-constraint-1} and \eqref{selfdual-constraint-2}.
This also shows that there are no totally antisymmetric tensors of rank 3
(for example, note that 
$[X^4,\varepsilon^{\mu\nu\rho\sigma}\cM_{\rho\sigma}X_\nu] = \varepsilon^{54\mu\nu\rho\sigma}\cM_{\rho\sigma}\cM_{4\nu} = 0$)

Since $X^\mu$ and $T^\mu$ are orthogonal via \eqref{TX-id} with radius related by \eqref{TT-XX-relation}, they can be reduced to 2 independent vector operators $X^i$ and $T^i$ under $SO(3)$. These generate the algebra of functions on an underlying 6-dimensional space, identified as twistor space $\C P^{1,2}$ in the following. 
Similarly, there are two $SO(4)$ vector operators $X^a$ and 
\begin{align}
\label{V-def}
V^a := \cM^{a0}, \qquad a=1,2,3,4,
\end{align}
which satisfy similar radius and orthogonality relations, notably 
\begin{align}
\label{X-V-contraction}
    X^a V_a + V_a X^a = 0 \ .
\end{align}

\item  {\bf antisymmetric rank 2 tensor operators}

These comprise $\cM^{\mu\nu}, X^\mu T^\nu - X^\nu T^\mu$, and $\tilde \cM^{\mu\nu} := \varepsilon^{\mu\nu\rho\sigma}\cM_{\rho\sigma}$. It turns out that  $\cM^{\mu\nu}$ can be expressed in terms of the $X^\mu$ and $T^\mu$ as follows
\begin{align}
\label{cM-X-T-id-1}
\cM^{\mu\nu} = c(X^4)(T^\mu X^\nu - T^\nu X^\mu) 
     + d(X^4) \varepsilon^{\mu\nu\rho\sigma} T_\rho X_\sigma \ .
\end{align}
To see this, we  observe that both sides commute with $X^4$:
\begin{align}
    [X^4,T^\mu X^\nu - T^\nu X^\mu] &=  [X^4,T^\mu ] X^\nu
    + T^\mu[X^4, X^\nu] 
    - [X^4,T^\nu ]X^\mu
    - T^\nu [X^4,X^\mu] \nn\\
    &= i/r [X^\mu, X^\nu]
    + i r^3 [T^\mu, T^\nu] =0 \ .
\end{align}
Diagonalizing $X^4$ reduces $\cH_0$ to irreps of $SO(3,1)$, and the claim follows. The  factors $c(X^4), d(X^4)$ are computed 
in the appendix \ref{sec:appendix-B}, where we show that  $d(X^4) = 0$ and
\begin{align}
\label{cM-X-T-id-2}
\boxed{
    \cM^{\mu\nu} =  r X_4^{-1}(T^\mu X^\nu - T^\nu X^\mu) \ .
      }
\end{align}
This simple form is specific for the $n=0$ case, while  for $n>0$ there is an extra term with $\varepsilon^{\mu\nu\rho\sigma}$ \cite{Sperling:2019xar} .

\item  {\bf symmetric tensor operators and functions in $\End(\cH_0)$}

Consider first 
the totally symmetric rank 2 tensor operators generated by $X^\mu$, which  are given by
 $\eta^{\mu\nu}$ and $X^\mu X^\nu + X^\nu X^\mu$.
Including also the $T$ generators, we have in addition $T^\mu X^\nu + T^\nu X^\mu$ and 
 $T^\mu T^\nu + T^\nu T^\mu$ at rank 2.
However, not all of them are independent due to the constraints \eqref{TX-id} and  \eqref{TT-XX-relation}
\begin{align}
 r^{-2} X_0^2 + r^2 T_0^2 &= r^{-2} X_i X^i + r^{2} T_i T^i \nn\\
 X_0 T_0 &=  X^i T_i + \frac {4i}r  X^4 \ 
\end{align}
which, together with \eqref{radius-constraint},
allows to determine $X^0,X^4$ and $T^0$ in terms of $X^i,T_j$  for $i,j=1,2,3$.
This means that all "functions" or operators in $\End(\cH_0)$ can be expressed as 
(ordered or totally symmetrized) polynomials or power series 
$\phi(X^i,T^j)$
in $X^i$ and $T^j$. Hence as vector space, $\End(\cH_0)$ can be identified with functions on $\R^6$.

\end{itemize}

We have thus recognized $\End(\cH_0)$ as quantized algebra of functions on a 6-dimensional space, which will be identified 
 in section \ref{sec:osc-CP12} as  twistor space $\C P^{1,2}$. 
 This space in turn will be viewed as a sphere bundle over space-time. The geometry will be made  explicit in terms of (quasi-) coherent states with good localization properties, which will
also allow to define a quantization map.

It is useful to note that the $SO(4)$ invariant subalgebra of $\End(\cH_0)$ is generated by $X^0$, while the $SO(3,1)$ invariant subalgebra is generated by $X^4$.
The $SO(3)$ invariant subalgebra  is generated by 
 $X^0, X^4, T^0$, and contains a 
 special central generator
    $X_4^2 - X_0^2 +  r^4 T_0^2$ that commutes with all $X^0, T^0, X^4$.

We summarize the commutation relations of 
minimal covariant quantum space-time\footnote{It is interesting to observe similarities with $\kappa$-Minkowski space \cite{Lukierski:1993wx}, if the extra generators $T^\mu$ were fixed. The $T^\mu$ allow to preserve a large symmetry group, which is the whole point of covariant quantum spaces.} $\cM^{1,3}_0$:
\begin{align}
\label{XX-CR-explicit}
\boxed{
\begin{aligned}
 [X^\mu,X^\nu] 
 &= -ir^3 X_4^{-1}(T^\mu X^\nu - T^\nu X^\mu) \\
    [T^\mu,X^\nu] &= \frac{i}{r} X^4\eta^{\mu\nu}\,   \\
     [T^\mu ,T^\nu] &= 
     \frac{i}{r} X_4^{-1}(T^\mu X^\nu - T^\nu X^\mu)  \\
    [X^4,X^\mu] 
    &= i r^3 T^\mu, \qquad
     [X^4,T^\mu] = \frac{i}{r} X^\mu\, .
\end{aligned}
}
\end{align}
Together with the constraints 
\eqref{constraints-CP12}, this provides a self-contained characterization of $\End(\cH_0)$.

\subsection{Oscillator construction and twistor space $\C P^{1,2}$}
\label{sec:osc-CP12}

It is useful to construct the  minimal doubleton representations $\cH_0$
using an oscillator construction.
Consider bosonic creation and annihilation operators $a_{i}, b_j$ which satisfy
\begin{align}
 [a_i, a_j^\dagger] = \delta_{i}^{j} \, , \qquad
[b_i, b^\dagger_j] = \delta_{i}^{j}  \qquad \text{for }i,j=1,2 \; .
\end{align}
Using these $a_{i}, b_j$, we define 4 operators  
\begin{align}
Z^\a := (a^\dagger_{1},  a^\dagger_{2}, b_1, b_2)
\end{align}
transforming as spinor of $SU(2,2)$,
with Dirac conjugates 
\begin{align}
\bar{Z} \equiv {Z}^{\dagger}\gamma^{0}
=
\left(-a_{1},-a_{2}, b^\dagger_{1}, b^\dagger_{2}\right)
\end{align}
transforming in the dual.
Here $\gamma^0 = \begin{pmatrix}
 -\one & 0 \\ 0 & \one
\end{pmatrix}$ is a gamma matrix associated to $SO(4,1)$  \cite{Sperling:2018xrm}.
They satisfy
\begin{align}
\label{Z-CR}
 [Z^\a,\bar Z_\beta] = \delta^\a_\beta \ ,
\end{align}
and it is an easy exercise to see that the
\begin{align}
 \label{oscillator-costruct-H4n}
 \cM^{ab} := \bar{Z}\Sigma^{ab} Z 
\end{align}
 satisfy the $\mso(4,2)$
commutation relations \eqref{M-M-relations-noncompact}, where the $\Sigma^{ab}$ are spinorial generators of $SU(2,2)$ with
 ${\Sigma^{ab}}^\dagger = \gamma^0 \Sigma^{ab} {\gamma^0}^{-1}$ \cite{Sperling:2018xrm}.
As a consequence, the $\cM^{\mu\nu}$ implement unitary representations of $SU(2,2)$.
Furthermore,
\begin{align}
 \hat N =  \bar{Z}Z = -N_a + N_b - 2 
\end{align}
commutes with the $SU(2,2)$ generators, where 
$N_{a} \equiv {a}^\dagger_{i} {a}_{i}$,
$N_{b} \equiv {b}^\dagger_{j} {b}_{j}$ are the standard bosonic number operators. Now
consider the Fock space
\begin{align}
 \cF = \oplus a^\dagger ... b^\dagger |0\rangle
\end{align}
spanned by the creation operators
${a}^\dagger_{i},{b}^\dagger_{j}$ acting on the Fock vacuum
$a_i \left|0\right\rangle = 0 = b_i \left|0 \right\rangle$.
This furnishes a unitary but reducible representation of $SU(2,2)$.
The minimal doubleton irrep is  obtained on the subspace
\begin{align}
\cH_0 \ := 
 \cF_{\hat N = -2} \qquad \mbox{i.e.} \quad N_a = N_b \ .
\end{align}
This yields a lowest-weight representation which is multiplicity-free, and therefore remains irreducible under $SO(4,1) \subset SO(4,2)$. It it is not hard to check that it satisfies all the relations in section \ref{sec:min-H0}.
Hence $\End(\cH_0)$ consists of (noncommutative) functions of 4 complex operators $Z_\a$ subject to $\bar Z Z = -2$ and invariant under $U(1)$. But this is precisely the characterization of twistor space\footnote{The relation with twistor space is discussed in more detail in \cite{Steinacker:2023ntw}.} $\C P^{1,2}$, and we conclude
\begin{align}
\label{End-twistor-space}
\boxed{ \
    \End(\cH_0) \cong \cC(\C P^{1,2}) \ .
\ }
\end{align}
This identification holds in the sense of almost-local functions \cite{Steinacker:2020nva}, i.e.
for semi-classical functions which are slowly varying on some scale $L_{NC}$.
In that regime, the canonical brackets \eqref{Z-CR} on $\C^4$
induce a Poisson (in fact a symplectic) structure on $\C P^{1,2}$ compatible with $SU(2,2)$, so that $\End(\cH_0)$ can be identified as quantized twistor space.
 This construction generalizes to the non-minimal siblings $H^4_n$, which provide a 1-parameter deformation of the underlying symplectic structure.
 
In particular, we note that the time-like generator $X^0$ is
given explicitly by
\begin{align}
 r^{-1}\, X^0 = \cM^{05} = \frac 12 (N_a + N_b + 2)
  = N_a + 1 \ .
\label{conf-energy}
\end{align}
The explicit form of the other generators in terms of the $a_i, b_j$ is given in appendix \ref{sec:hilbertspace}.

\section{4-dimensional geometry arising from $\cH_0$}

The above construction allows to 
identify $\End(\cH_0)$  with the space of functions on twistor space $\C P^{1,2}$.
A priori, an algebra of functions does not contain any metric structure. A geometric interpretation is obtained by considering some matrices as quantized embedding functions of the underlying symplectic space $\cM$ in a
target space  with  flat metric. 
One possibility is to consider all $\msu(2,2)$ generators
\begin{align}
    \cM^{ab} \sim m^{ab} : \quad \cM \hookrightarrow \msu(2,2) \cong \R^{15}
\end{align}
which allows to identify $\cM \cong \C P^{1,2}$.
Alternatively, we can choose some subset of these generators, e.g.
\begin{align}
    X^a \sim x^a : \quad \cM \hookrightarrow \R^{1,4}
\end{align}
for $a=0,...,4$. This can be viewed as a projection of $\C P^{1,2}$ to $\R^{1,4}$, and
 the radial constraint suggests an interpretation in terms of fuzzy $H^4 \subset \R^{1,4}$.
Other choices of embedding functions yield different pictures or projections, some of which are discussed in the following.
The most "appropriate" picture depends on the 
matrix configuration 
realized in the matrix model.

\subsection{Projection to $\R^4$}
\label{sec:R-4-H}

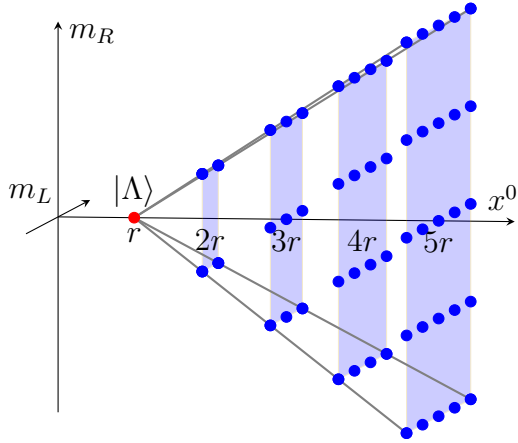
\begin{figure}[h!]
\begin{center}
\begin{tikzpicture}
    \begin{axis}[
        view={8}{5}, 
        xlabel={$x^0$},
        ylabel={$m_L$},
        zlabel={$m_R$},
        grid=both,
        axis lines=middle,
        xmin=0, xmax=6,
        ymin=-4, ymax=4,
        zmin=-4, zmax=4,
        ticks=none
    ]
    
    \newcommand{\drawlayer}[2]{%
        \pgfmathsetmacro{\halfsize}{#2 - 1}
        \pgfmathsetmacro{\step}{2 * \halfsize / (#2 - 1)}
        \pgfmathsetmacro{\numDivisions}{#2 - 2} 

        \foreach \i in {0,...,#2} {%
            \ifnum\i<#2 
                \pgfmathsetmacro{\pos}{-\halfsize + \i * \step}
                \addplot3[only marks, mark=*, color=blue] coordinates {(#2, \pos, -\halfsize)};
                \addplot3[only marks, mark=*, color=blue] coordinates {(#2, \pos, \halfsize)};
                \addplot3[only marks, mark=*, color=blue] coordinates {(#2, -\halfsize, \pos)};
                \addplot3[only marks, mark=*, color=blue] coordinates {(#2, \halfsize, \pos)};
            \fi
            
        }
        \foreach \i in {1,...,\numDivisions} { 
            \foreach \j in {1,...,\numDivisions} { 
                \pgfmathsetmacro{\posX}{-\halfsize + \i * \step}
                \pgfmathsetmacro{\posY}{-\halfsize + \j * \step}
                \addplot3[only marks, mark=*, color=blue] coordinates {(#2, \posX, \posY)};
            }
        }

        \addplot3[patch, patch type=rectangle, fill=blue, opacity=0.2] coordinates {
            (#2, -\halfsize, -\halfsize)
            (#2, -\halfsize,  \halfsize)
            (#2,  \halfsize,  \halfsize)
            (#2,  \halfsize, -\halfsize)
        };
    }

    \newcommand{\connectlayers}[2]{%
        \pgfmathsetmacro{\halfsizeA}{#1 - 1}
        \pgfmathsetmacro{\halfsizeB}{#2 - 1}
        \addplot3[thick, color=gray] coordinates {(#1, -\halfsizeA, -\halfsizeA) (#2, -\halfsizeB, -\halfsizeB)};
        \addplot3[thick, color=gray] coordinates {(#1, -\halfsizeA,  \halfsizeA) (#2, -\halfsizeB,  \halfsizeB)};
        \addplot3[thick, color=gray] coordinates {(#1,  \halfsizeA, -\halfsizeA) (#2,  \halfsizeB, -\halfsizeB)};
        \addplot3[thick, color=gray] coordinates {(#1,  \halfsizeA,  \halfsizeA) (#2,  \halfsizeB,  \halfsizeB)};
    }

    \addplot3[only marks, mark=*, color=red] coordinates {(1, 0, 0)}; 
    \node at (axis cs:1, 0, 0) [anchor=south] {$\ket{\Lambda}$}; 
    
    \drawlayer{2}{2} 
    \node at (axis cs:2, 0, -1) [anchor=south] {}; 
    
    \drawlayer{3}{3} 
    \drawlayer{4}{4} 
    \drawlayer{5}{5} 

    \connectlayers{2}{3}
    \connectlayers{3}{4}
    \connectlayers{4}{5}

    \pgfmathsetmacro{\halfsizeTwo}{2 - 1}
    \addplot3[thick, color=gray] coordinates {(1, 0, 0) (2, -\halfsizeTwo, -\halfsizeTwo)};
    \addplot3[thick, color=gray] coordinates {(1, 0, 0) (2, -\halfsizeTwo,  \halfsizeTwo)};
    \addplot3[thick, color=gray] coordinates {(1, 0, 0) (2,  \halfsizeTwo, -\halfsizeTwo)};
    \addplot3[thick, color=gray] coordinates {(1, 0, 0) (2,  \halfsizeTwo,  \halfsizeTwo)};
    \node at (axis cs:1, 0, 0) [anchor=north] {$r$};
    \node at (axis cs:2, 0, 0) [anchor=north] {$2r$};
    \node at (axis cs:3, 0, 0) [anchor=north] {$3r$};
    \node at (axis cs:4, 0, 0) [anchor=north] {$4r$};
    \node at (axis cs:5, 0, 0) [anchor=north] {$5r$};

    \end{axis}\label{weights}
\end{tikzpicture}
\end{center}
\caption{Weight structure of the minimal doubleton representation $\cH_0$. All mutliplicities are 1.}
\label{fig:weights-minirep}
\end{figure}

Consider the 4 generators $X^i, \ i=1,2,3,4$ which transform as vectors of $SO(4)$.
The matrix configuration defined by these 4 matrices
defines a covariant 4-dimensional Euclidean space $\R^4$. 
The Hilbert space $\cH_0$ decomposes accordingly into the direct sum of subspaces\footnote{Incidentally, this decomposition allows to identify $\cH_0$ with the space of (commutative) scalar functions on $S^3 \cong SU(2)$  via the Peter-Weyl theorem, as a bi-module of $SU(2)_L \times SU(2)_R$. This arises from the $SO(4,1)$ group action as conformal transformations of $S^3$.}
\begin{align}
\label{H-0m-decomp}
    \cH_0 = \bigoplus_{m=0}^\infty 
 \cH_{0,m} 
\end{align}
where 
\begin{align}
    \cH_{0,m} =  (m)_L \otimes (m)_R = \{X^0 = R_m\}, \qquad R_m = (m+1)r
\end{align}
and $(m)_{L,R}$ denotes $m$-dimensional irreps of $SU(2)_L$ and $SU(2)_R$, respectively, as shown in figure \ref{fig:weights-minirep}.
Clearly all $\cH_{0,m}$
are preserved by $SO(4)$, which is generated by $\cM^{ij}$ and $r T^i = \cM^{i4}$; in contrast, the non-compact generators in 
 $SO(4,1)$ link different  $\cH_{0,m}$. 
Moreover, the radius
\begin{align}
X \cdot X \equiv\sum\limits_{i=1}^4 X_i^2  = r^2 + X_0^2 = r^2 + R_m^2
\end{align}
is constant on each $\cH_{0,m}$ using  \eqref{radius-constraint} and measured by $X^0$. 
This suggests to interpret $\cH_{0,m}$ in terms of a fuzzy 3-sphere $S^3$ with radius $R_m$;
however this is problematic, because the $X_i$ generators do not respect $\cH_{0,m}$. It makes more sense to associate 
a 4-dimensional ball to the direct sum of the $\cH_{0,m}$ up to a radius $R$:
\begin{align}
\label{B4N}
    \B^4_{N} := \bigoplus_{m=0}^N \,\cH_{0,m}
    \quad   
    \cong \  \{X \cdot X \leq R_N^2\}, \qquad R_N = (N+1) r \ .
 \end{align}
 The space of operators on this subspace clearly describes some sort of 4-dimensional Euclidean ball  of radius $R_N$; however, there is a hidden $S^2$ fiber\footnote{This reflects the fact that the underlying symplectic space is 6-dimensional.}, which can be described by the extra generators $V_a$ \eqref{V-def}.

It is interesting to compute the  number of states associated with the ball $\B^4_N$  of radius $R \sim N r$.
Using $\dim\cH_{0,m} = (m+1)^2$, we obtain
\begin{align}
\label{cH-R-dim}
    \dim (\B^4_N) 
     = \sum_{m=0}^N m^2 
     \sim \frac 13 N^3 
     \sim 
    \frac 1{3} \Big(\frac {R_N}r\Big)^3 
    \sim {\rm A} (S^3_R = \partial B_R^4)
\end{align}
for large $N$.
This is proportional to the area of the 3-sphere at the boundary, rather than the volume of the enclosed space. A different interpretation of that result will be found below.

\subsection{Projection to  $\cM^{1,3}  \subset \R^{1,3}$}
\label{sec:M13-H}

\begin{figure}[h!]
    \centering
    \includegraphics[width=.7\linewidth]{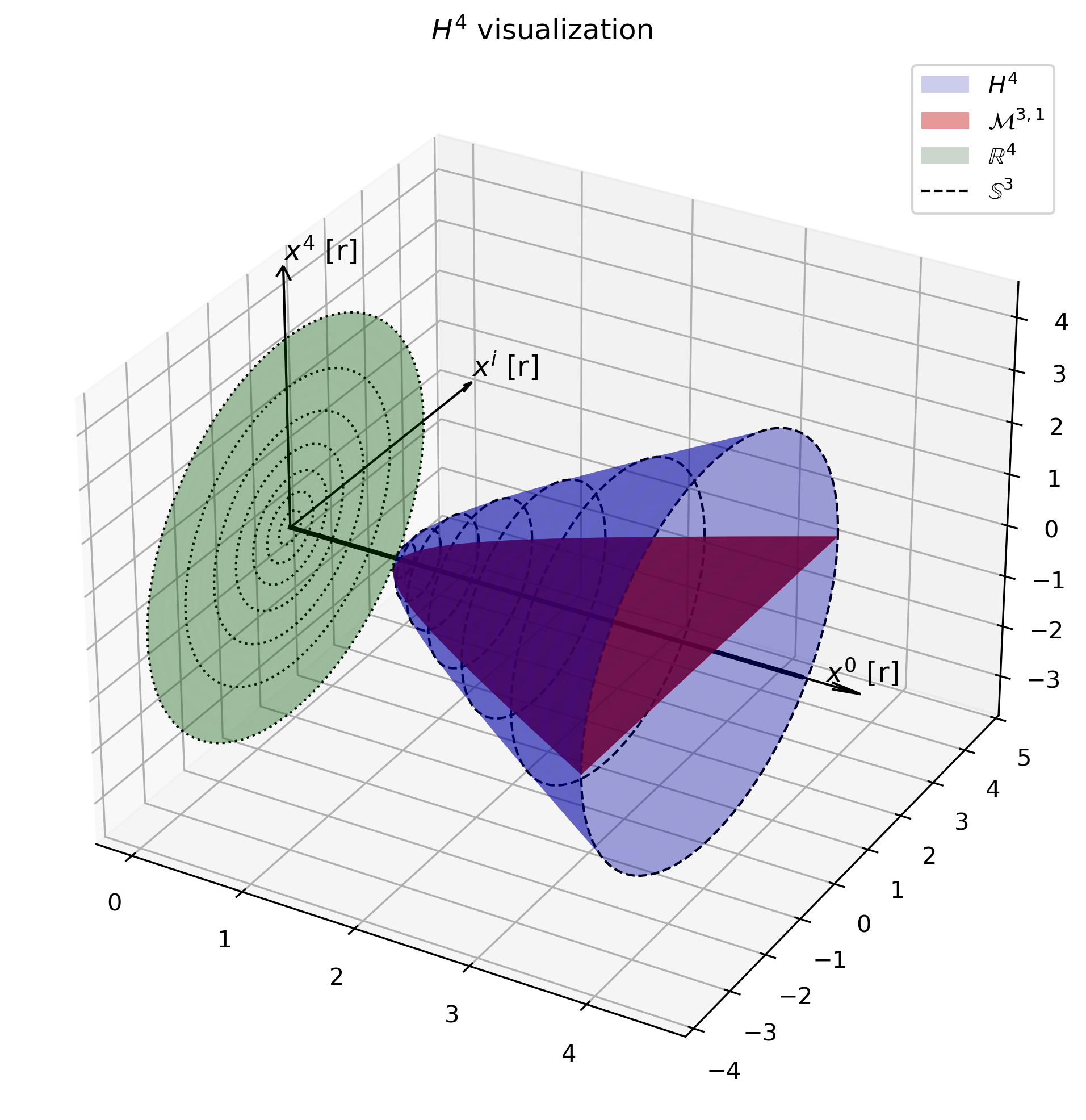}
    \caption{Lower dimensional visualization of $H^4 \subset \R^{1,4}$ and the projections to $\mathcal{M}^{3,1}$ and $\R^4$, indicating also the concentric spheres $S^3$.}
    \label{hyT_proj}
\end{figure}

\begin{figure}[h!]
    \centering
    \includegraphics[width=.7\linewidth]{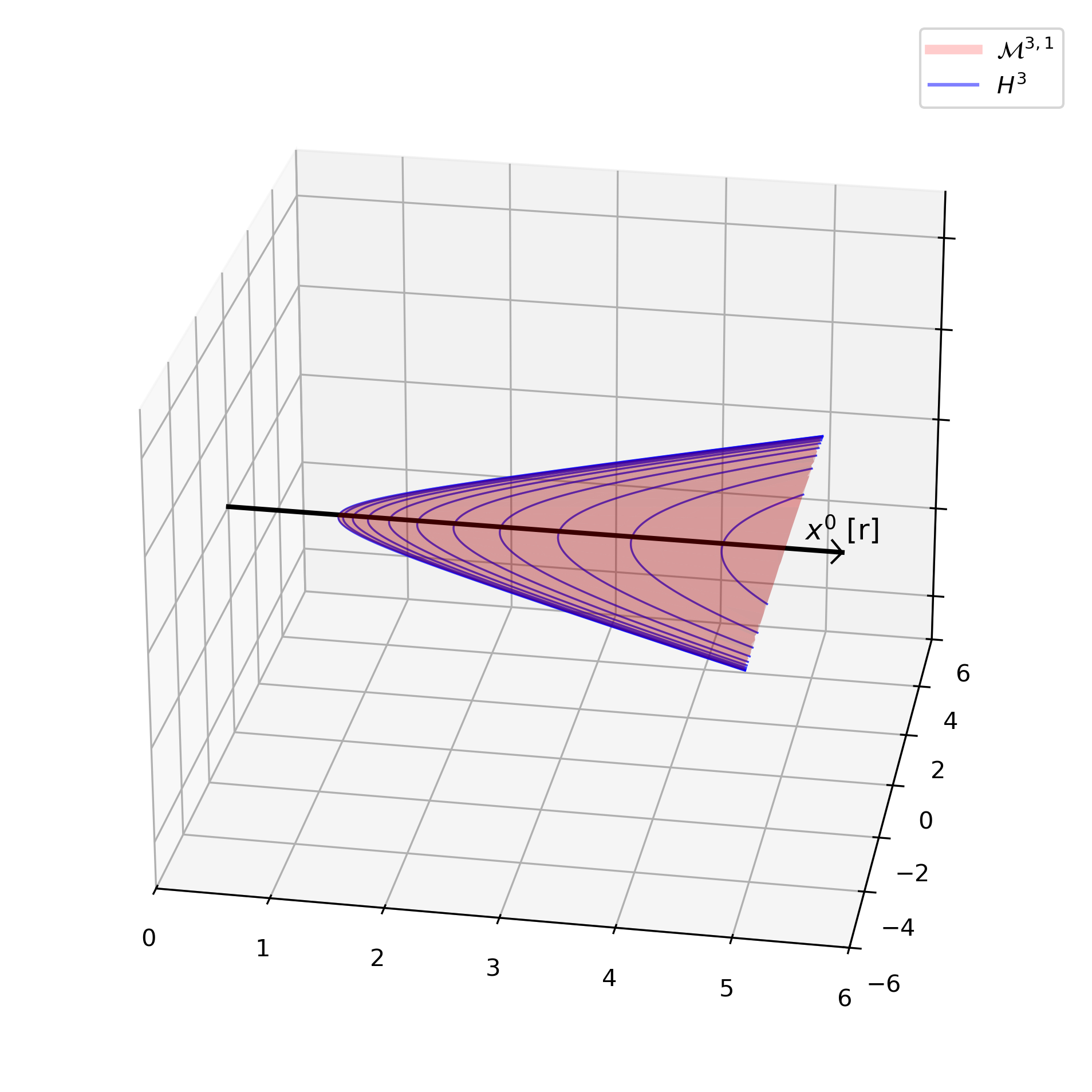}
    \caption{Visualization of $\mathcal{M}^{3,1}$ and its foliation into space-like hyperboloids $H^3$ defined by $\tau$.}
    \label{h3foliation}
\end{figure}

In the same vein, we can consider the embedding defined by the 3+1 matrices 
\begin{align}
    X^\mu \sim x^\mu: \quad \cM \hookrightarrow \R^{1,3}_x
\end{align}
transforming as vectors of $SO(3,1)$
for $\mu=0,...,3$. This can be viewed as a projection of $\C P^{1,2}$ or $H^4$ to $\R^{1,3}$. These different projections are visualized in figure \ref{hyT_proj}.
Now $\cH_0$ decomposes into a direct integral of unitary  $SO(3,1)$ irreps according to the the eigenvalues of $X_4$ according to \eqref{SO31-casimir-1}, interpreted as square-integrable scalar functions on $H^3$.
This picture  leads to an interpretation in terms of a 3+1-dimensional  space-time denoted as  $\cM^{1,3}$ equipped with Lorentzian metric,
 foliated into hyperbolic space-like $H^3$ \cite{Sperling:2019xar}.
 In the framework of matrix models,
 that metric turns out to describe a cosmological $k=-1$ FLRW space-time which is asymptotically flat at late times $ r^{-1} x^0 \gg 1$, featuring a Big Bounce at the fold connecting
the upper and lower sheets\footnote{It is also possible to obtain a $k=0$ FLRW space-time with Big Bang, with yet another projection \cite{Steinacker:2024unq}.}.
Moreover there is a hidden $S^2$ fiber, which is not resolved by the above projection. That fiber is resolved by the extra 3+1 matrices 
\begin{align}
    T^\mu \sim t^\mu: \quad \cM \hookrightarrow \R^{1,3}_p \ .
\end{align}
In the {\bf semi-classical regime}, the constraints \eqref{radius-constraint}, 
\eqref{TX-id}
reduce to\footnote{For the non-minimal doubletons $\cH_n$ with $n>0$, the first constraint would read $x_a x^a \approx - \frac{n^2-4}4 r^2$.}
\begin{align}
\label{constraints-semiclass}
    x_a x^a &= r^2 \approx 0,    \nn\\
    \qquad x_\mu t^\mu &= 0 \ .
\end{align} 
The $r^2$ on the rhs requires some discussion.
Since the spectrum of $X^0$ is positive \eqref{spec-X0}, the $X^a \sim x^a$ describe a one-sided hyperboloid as in figure \ref{hyT_proj},
corresponding to $x_a x^a \approx - r^2 \approx 0$, which approximately coincides with the light-cone $x^a x_a \approx 0$ at late times. Hence the "wrong sign" on the rhs of \eqref{constraints-semiclass} is a small quantum effect, as discussed below.
Accordingly, we will parametrize 
the semi-classical hyperboloid  with hyperbolic coordinates as follows
\begin{align}
    \begin{pmatrix}
     x^0\\
     x^1\\
     x^2\\
     x^3
    \end{pmatrix}
    &=r\cosh(\tau)\begin{pmatrix}
     \cosh(\chi)\\
\sinh(\chi)\sin(\theta)\cos(\varphi)\\
\sinh(\chi)\sin(\theta)\sin(\varphi)\\
     \sinh(\chi)\cos(\theta)
    \end{pmatrix}\,  \nn\\[1ex]
    x^4 &= r \sinh(\tau) \ .
\end{align}
Here $\tau$ can be recognized as a time-like parameter
of the resulting FLRW geometry on space-time $\cM^{1,3}$ \cite{Steinacker:2017bhb,Sperling:2019xar}, foliated into space-like hyperboloids $H^3$
defined by $x^4 = r \sinh(\tau) = const$ 
as in figure \ref{h3foliation}.

By considering a local patch $x^\mu \approx\xi^\mu$ around a reference point
\begin{align}
\label{ref-point}
    \xi^\mu = (\xi^0,0,0,0) \quad \in \cM^{1,3}
\end{align}
the constraint $x_\mu t^\mu = 0$ implies that $t^0 \approx 0$, so that the $t^\mu$ generators describe a space-like 2-sphere $S^2$ with radius $r^{-2} \xi^0$. 
Near this reference point, the commutators \eqref{XX-CR-explicit} the reduce to Poisson brackets
 \begin{align}
 \label{Poisson-brackets}
     \{x^i,x^0\} &= -\frac{r^2 x^0}{\sinh(\t)} t^i  
     \qquad \approx -ir^3 t^i \nn\\
    \{x^i,x^j\} &= -\frac{r^2}{\sinh(\t)}(t^i x^j - t^j x^i) \approx 0 \ .
\end{align}
 This provides a local description of the 6-dimensional twistor space $\C P^{1,2}$ as a sphere bundle over space-time.
 We will justify this semi-classical picture in section \ref{sec:6D-states} by constructing quasi-coherent states with good localization properties on the 6-dimensional bundle. 
In particular, the 4-ball $\B^4_N$ \eqref{B4N} can now be understood as a sector of the cosmic FLRW space-time $\cM^{3,1}$ bounded by some compact space-like ("partial-Cauchy") surface 
\begin{align}
 \label{Cauchy-surface-x0}
    S^3_{R} := \{x^0 = R\}  \ \subset  \cM^{1,3}
 \end{align}
 on both sides of the Big Bounce.

It should be clear that despite these different possible interpretations,
the algebra of functions is always the same, and different pictures lead to  complementary insights.
Both pictures can be understood as different projections of the hyperboloid $H^4$ described by the $X^a$, $a=0,...,4$, and the  localized states constructed in section \ref{sec:6D-states} will apply in either picture, for large $x^0$. 
Using yet another projection (cf. section 5.5 in \cite{Steinacker:2024unq}), it is also possible to obtain a $k=0$ FLRW space-time with a Big Bang, which will be discussed in more detail elsewhere.

\subsection{Remarks on Early Universe}

The above semi-classical description applies for the "late time" regime\footnote{Strictly speaking, the considerations in this paper apply for the regime $r^{-1} x^0\gg 1$. This can be extended to $r^{-1}x_4\gg 1$ using $SO(4,1)$, which characterizes the late-time regime in the cosmic space-time.} $\frac{x_4}{r}\gg 1$. Let us briefly discuss the issues with the early time regime.

From the weight structure of the representation shown in figure \ref{weights}, we notice that to small $X^0$ eigenvalues correspond very few degrees of freedom at fixed $x^0 = O(r)$.
In that regime, the  uncertainty of the
noncommuting matrices is of order $r$, 
comparable with the radius of the  $S^3$. Therefore the semi-classical geometric interpretation fails. 
We can see this also from the foliation into 3-hyperboloids in figure \ref{hyT_proj}, which for $x_4\approx 0$, develops a "singularity". 
In the same vein, the "wrong sign" of the matrix constraint
\begin{equation}
    X^a X_a=+r^2\approx 0
\end{equation}
for $n=0$ arises because the quantum fluctuations
of the matrices become larger than the expectation values around $x^4 = O(r)$.
This means that the geometric description simply fails near the origin, or more precisely near the Big Bounce.
That, of course, is entirely reasonably physically,
and the near Big Bounce physics is described by matrices with no semi-classical counterpart.
In the non-minimal $H^4_n$ case with $n\gg 1$, the semi-classical notion of geometry does apply for any $\tau$.

\section{Space of functions and trace}
\label{sec:trace}

Consider a generic operator in $\End(\cH_0)$ expanded in powers of $T^\mu$
\begin{align}
\label{phi-hs-expansion}
    \Phi(X,T) &= \Phi_0(X) + \Phi_\mu(X)T^\mu + \Phi_{\mu\nu}(X) T^\mu T^\nu + ...
    = \sum_{s} \Phi_{\und{\mu}}(X) T^{\und{\mu}}   \nn\\
    &\in  \quad \cC^0 \quad \oplus  \quad \ \cC^1 \quad 
 \oplus ...
\end{align}
The $\Phi_{\und{\mu}}(X)$ are viewed as $\hs$-valued spin $s$ tensor fields on $\cM^{1,3}$, which can be assumed to be traceless i.e. all contractions with $\eta^{\mu\nu}$ vanish.
As shown in \cite{Sperling:2018xrm}, these
 "higher-spin sectors" $\cC^s$ can be defined as eigenspaces of a $\mso(4,1)$ Casimir $\cS^2 = 2s(s+1)$.

\paragraph{Trace and integration over $\R^4$.}

Using the decomposition \eqref{H-0m-decomp} of $\cH_{0}$ into irreducible $SO(4)$ modules, the trace can be written as a sum of the traces $\tr_m$ over the $\cH_{0,m}$
\begin{align}
    \tr(\Phi) = \sum_m \tr_m (\Phi) \ .
\end{align}
This can be interpreted as an integral  at fixed time $x^0$, followed by an integration over discrete time. 
The trace is of course invariant under the full $SO(4,2)$, since
\begin{align}
   0 &=  \tr [\cM^{ab},\Phi] 
\end{align}
for $\Phi \in \End(\cH_0)$ 
with good decay properties for large $X^0$.

In general, the dimension of the Hilbert space associated to a ball 
$B^4_R = \{X \cdot X \leq R^2\}$
is expected to coincide with its volume, as 
 defined by an underlying symplectic volume form. 
More generally, the trace over some observable should reproduce the integral over the corresponding semi-classical function.
To find the underlying semi-classical volume form, it is enough to consider functions $\Phi(X^0)$. Then 
\begin{align}
    \tr(\Phi) &= \sum_m (m+1)^2 \Phi(m+1) =  \sum_m \frac 1{m+1} (m+1)^3 \Phi(m+1) \nn\\
     &\sim \frac 1{2\pi^2 r^3}\int dx^1 ... dx^4\, \frac 1{x^0} \Phi(x^0) 
    = \int \Omega\, \Phi(x^0)
\end{align}
using ${\rm vol} (S^3_R) = 2 \pi^2 R^3$
for 
\begin{align}
\label{Omega-exact}
    \Omega = \frac{1}{2\pi^2} \frac{dx^1 ... dx^4}{r^3 x^0} \, .
\end{align}
In particular, we recover 
\begin{align}
\int\limits_{B_R^4} \Omega 
= \frac 1{r^3}\int_0^{R} \frac{R'^3 dR'}{\sqrt{r^2 + R'^2}} 
 \sim \frac{1}{3} \Big(\frac Rr\Big)^3
  =  \dim (\B^4_N)
\end{align}
 using $x^0 = \sqrt{r^2 + \sum_i x_i^2}$
for  $r^{-1}x^0 \gg 1$,
in agreement with \eqref{cH-R-dim} for large $R$. 
This uniquely determines the $SO(4,1)$-invariant  $\Omega$, at least at late times.
In the context of matrix models \cite{Steinacker:2019fcb,Steinacker:2024unq}, the symplectic volume form $\Omega = \rho_M d^4 x$ can be related to the 
Riemannian (metric) volume form via $\sqrt{G} \sim 
\rho_M\rho^{2}$, where $\rho$ is the dilaton.

It is worth noting that the above derivation generalizes easily to non-minimal doubletons $\cH_n$ with $n>0$, where 
the Hilbert space $\cH_n$ decomposes accordingly into the direct sum of subspaces
\begin{align}
    \cH_n = \oplus_m 
 \cH_{n,m} = \bigoplus_{m\geq 0} \, (m+n)_L \otimes (m)_R \ .
\end{align}
Now $\dim\cH_{n,m} = m(m+n)$. Hence for the 4-dimensional ball of radius $R^2 \sim r^2 N^2$, the dimension of the Hilbert space
\begin{align}
\label{cHn-R-dim}
    \dim (\B_N^4) 
    = \sum_m m(m+n) 
    \sim 
    \frac 1{3} \Big(\frac Rr\Big)^3  = \int_{B_R^4} \Omega
\end{align}
 for large $N$  at late times $\frac{m}{n} \gg 1$, with the same
symplectic form \eqref{Omega-exact} for any $n$. 
In that sense, the different quantum spaces $\C P^{1,2}$ arising from $\cH_n$ are very similar\footnote{In particular there is no extra factor $n$
in the volume form, correcting an error in 
\cite{Steinacker:2024huv}}.

\paragraph{Integration over space-time $\cM^{1,3}$.}

Consider now  modes organized in the form \eqref{phi-hs-expansion}, 
with a cutoff in $s$. 
We have seen that the trace can be computed by first projecting to the singlet sector in $SO(4)$ which leads to a function in $X^0$.
From the point of view of space-time, it is more natural to project first to the $\cC^0$ sector, since the trace vanishes for $s>0$. 
This can be achieved by first projecting the polynomials in $T_\mu$ to 
polynomials in $X^\mu$ in an 
$SO(3,1)$ invariant way respecting the constraints, which is 
achieved by
\begin{align}
\label{kappa-def}
    [T^\mu T^\nu]_0 = \frac 1{3r^2}\kappa^{\mu\nu} , \qquad 
    \kappa^{\mu\nu} = \frac 1{r^2}\big(\eta^{\mu\nu} (X_4^2-r^2) + X^\mu X^\nu\big)
\end{align} 
and suitable generalizations at higher order, cf \cite{Steinacker:2019awe}.
In the semi-classical regime, this 
projection 
$[t ... t]_0$  to $\cC^0$
amounts to integration  over the internal $S^2$ fiber described by $t^\mu$.
Then the integral is obtained  by integrating the remaining function of $x$ over space-time  
\begin{align}
    \tr(\Phi) \sim \int\limits_{\cM^{1,3}}\Omega [\Phi]_0 
\end{align}
with the volume form \eqref{Omega-exact}, which can be rewritten using the radial constraint \eqref{constraints-semiclass}  as 
\begin{align}
\label{Omega-exact-E-M}
   \Omega = \frac{1}{2\pi^2 r^3} \frac{dx^1 ... dx^4}{x^0} \,
 =  \frac{1}{2\pi^2 r^3} \frac{dx^0 ... dx^3}{x^4} \ .
\end{align}

  \section{Quasi-coherent states on twistor space $\C P^{1,2}$}
\label{sec:6D-states}
In this section we will construct appropriate quasi-coherent states that probe the 6-dimensional geometry of quantized twistor space $\C P^{1,2}$.\\

The underlying 6-dimensional space can be seen by recalling the commutation relations\footnote{Recall that $X^0$ and $T^0$ are functions of the space-like generators.}
\begin{align}
    [X^i,T_j] &= \frac{X^4}{r} \delta^i_j  
\end{align}
while $[X^i,X^j]$ and $[T^i,T^j]$ are much smaller, given by
\begin{align}
 [X^i,X^j] &= \frac{r^3}{X^4} (T^i X^j - T^j X^i) 
 = - r^2 [T^i,T^j] \ \ll \ [X,T]
\end{align}
due to \eqref{XX-CR-explicit}.
Hence locally for $X^4 \approx const$, the space is similar to canonical commutation relations $\R^6_\theta$.
Accordingly, we should find optimally localized quasi-coherent states which capture this 6-dimensional space.

One way to address this 
is by considering the matrix configuration $(X^i,T^i)$ for $i=1,2,3$ as background and computing the resulting quasi-coherent states numerically; this 
will be discussed in section \ref{sec:quasicoh-num}.
However, we first discuss an analytical approach which provides more insights.

\subsection{Analytical approach from oscillator construction}
\label{sec:coh-states-osc}

An analytically accessible construction of such (quasi-) coherent states is obtained from the oscillator construction of $\cH_0$ discussed in section \ref{sec:osc-CP12}, based on
the standard (shifted) coherent states on $\C^4_\theta$.
Consider canonical coherent states 
located at some large expectation value $\langle a_i\rangle = \a_i$ and $\langle b_i\rangle = \beta_i$. 
Projecting to $\cH_0 \ = \cF_{\hat N = -2}$  with $N_a = N_b$ should give  states in $\cH_0$ which are well localized in 6 dimensions, and
 can be used for analytic computations.

Explicitly, consider a canonical coherent state in the full Fock space:
\begin{align}
    |\a,\b\rangle_\cF = U_\a U_\b |0,0\rangle_\cF \quad \in \cF
\end{align}
located at any $(\a_i,\b_j) \in \C^4$, where $U_a$ are translation operators on the quantum planes $\C^2_\theta$ generated by $a_i$ and $b_j$. 
For example, we can choose coherent states located at  
\begin{align}
\label{location-C4}
\b_i= \begin{pmatrix}\sqrt{m}\, e^{i\phi}\\ 0 \end{pmatrix}, \quad
\a_i= \begin{pmatrix}\sqrt{m} \\ 0\end{pmatrix} \ \ \in \ \C^2 \times \C^2 \ .
\end{align}
Now consider their projection
\begin{align}
   |\a,\b\rangle :=  \Pi_{0} |\a,\b\rangle_\cF
\end{align}
to $\cH_0$ with $N_b = N_a$. These are the desired localized states on $\C P^{1,2}$.
We can estimate the expectation values of the generators in the semi-classical regime (i.e. $m \gg 1$) as follows:
\begin{align}
r^{-1}x^0 \approx
    r^{-1}\langle X^0\rangle = \frac 12(N_a + N_b + 2) = N_a+1 = m+1\ 
\end{align}
using $\hat N=-2$. 
That constraint implies  
\begin{align}
 r^{-1} x^0  \approx  \a^\dagger \a \approx \b^\dagger \b \approx m
\end{align}
which means that $\a_i$ and $\b_i$ are of order $\sqrt{x^0/r}$.
We then obtain
using 
 \eqref{osc-ops-explicit}
 \begin{align}
    \langle X^i\rangle &\approx \frac r2(\a^t \sigma^i \b + \b^\dagger\sigma^i \a^*) \approx \ (0,0,\oveq{\bar x^3}{rm\cos\phi}) \nn\\
     \langle X^4\rangle  &\approx \frac {i r}2(- \a_j \b_j + \b_j^* \a_j^*)   \approx rm\sin\phi=:\bar x^4 , \nn\\
     \langle X^0\rangle &\approx r m=:\bar x^0 ,  \nn\\
     \langle T^i\rangle &\approx \frac 1{2r}(\a^\dagger \sigma^i \a + \b^\dagger \sigma^i \b) \ \approx \ (0,0,\oveq{\bar t^3}{r^{-1}m})\nn\\
     \langle T^0\rangle &\approx 
      \frac 1{2r}(\a_j \b_j + \b_j ^* \a_j^*) \ 
  \approx r^{-1} m\cos\phi=:\bar t^0 \ .
     \label{location-x-t-a-b}
\end{align}
Therefore the state is localized at 
\begin{align}
  x^a=\langle X^a \rangle =  (\bar x^0, 0, 0, \bar x^3, \bar x^4) = \bar x^a,
  \qquad  t^\mu=\langle T^\mu \rangle = (\bar t^0,0,0,\bar t^3)=\bar t^\mu
\end{align}
which satisfies the constraints 
\begin{align}
\label{constraints-approx}
    x_a  x^a \approx 0 \approx x_\mu t^\mu
\end{align}
consistent with \eqref{constraints-semiclass} up to uncertainties.
 Since the uncertainty of canonical coherent states is given by $\Delta a^i = 1$,  the uncertainty of $X^a$ and $T_\mu$ is of order 
\begin{align}
    \Delta X^\mu &\sim  r \sqrt{m}   
 \ = \  O(L_{\rm NC}) ,   \nn\\
    \Delta T^\mu &\sim   r^{-1} \sqrt{m}  \ = \   O(r^{-2} L_{\rm NC}) \ .  
\end{align}
Here 
the scale of noncommutativity is given by
\begin{align}
\label{LNC}
    L_{\rm NC} = r \sqrt{m} 
    = r \sqrt{\frac{\bar x^0}{r}} \ \gg r 
\end{align}
as expected from \eqref{Poisson-brackets}.  
All these considerations are of course unchanged under arbitrary
$SU(2)_L \times SU(2)_R$ rotations, which preserve $x^0$ and allow to  choose the location of these canonical coherent states on $\C^4_\theta$ such that the expectation values $\langle X^\mu\rangle$ and $\langle T_\mu\rangle$ recover any point on $\C P^{1,2}$.
We will accordingly denote them as $\ket{x,t}$, with
\begin{align}
\label{exp-values-coh-analytical}
    \bra{x,t} X^a \ket{x,t} = x^a,
  \qquad \bra{x,t} T^\mu \ket{x,t} =  t^\mu
\end{align}
sweeping out $\C P^{1,2}$ consistent with \eqref{constraints-semiclass}.
In particular, the uncertainty of $\Delta T^\mu$ implies that the internal $S^2$ supports only 
$O(m)$ quantum cells, and can thus be viewed as a sort of fuzzy sphere. This implies that the 
number of higher-spin $\hs$ modes at some given point in space-time is limited by 
\begin{align}
\label{hs-cutoff}
  s \leq m = r^{-1} x^0 \ .
\end{align}
This is important for the UV finiteness of the (local part of the) one-loop effective action.

The above coherent states can also be understood in terms of the two fuzzy spheres generated by the angular momentum generators $L^i_{L,R}$ defined by the $a_i$ and $b_j$ generators respectively, as in section \ref{sec:fuzzy-S2-coh}. 
In particular, $T^i$ generates the "diagonal sum" of these two fuzzy spheres $S^2_m$ with  $m = \frac{x^0}{r}$, and we recover the above  cutoff for the $\hs$ spectrum.

These considerations are easily adapted to fuzzy $\C P^{1,2}_n$ for any $n\geq 0$. The main difference is a modified radius constraint, which however becomes irrelevant in the late-time regime. Therefore all these spaces essentially coincide at late times, and we can restrict ourselves to the minimal case $\cH_0$.

\subsection{Completeness relation and quantization map}

It is instructive to consider the orbit 
of the canonical coherent states $|\a,\b\rangle_\cF$ under $SU(2,2)$.
These states are closed under the compact $SO(4) \times U(1)$, but not under the noncompact transformations; the latter would create squeezed states different from the canonical ones.
 Note that $SO(4)$ acts transitively on the space-like sphere $S^3_R$ \eqref{Cauchy-surface-x0}
 for fixed $\t$, and the orbit is recognized as an $S^2$ bundle 
\begin{align}
 \label{Cauchy-surface-x0-full}
    \tilde S^3_R := \{x^0 = R\} \stackrel{loc}{\cong} S^3 \times S^2 \quad \cong SO(4)/U(1) 
 \end{align}
 over the space-like $S^3_R$.
 Therefore the $|\a,\b\rangle$ states fully capture the bundle geometry, maintaining the full $SO(4)$ invariance in the NC case.
These states can hence be used for local QFT calculations, including the $\hs$ modes along the lines of \cite{Steinacker:2024huv,Steinacker:2023myp}, and to construct string modes in section \ref{sec:string-modes}.
In particular,
the completeness relation on $\C^4_\theta$
\begin{align}
    \int_{\C^4}  |\a,\b\rangle_\cF \langle \a,\b|_\cF = c\one_\cF \ 
\end{align}
implies by projecting to $\cH_0$ 
\begin{align}
   \one_{\cH_0} =  \int\limits_{\C P^{1,2}} \tilde\Omega |\a,\b\rangle \langle \a,\b| 
  = \int\limits_{R^4} \Omega\int\limits_{S^2} |x,t\rangle \langle x,t|
   =  \int\limits_{\cM^{1,3}} \Omega\int\limits_{S^2}|x,t\rangle \langle x,t|
\end{align}
for some $SO(4)$-invariant measures $\tilde \Omega$ and $\Omega$.
The integral is over the subspace $|\a|^2 = |\b|^2$ of $\C^4$, which is a $U(1)$ bundle over $\C P^{1,2}$.
These formulas are equivalent to the trace formulas   
\begin{align}
    \tr(\Phi) = \int\limits_{R^4} \Omega\int\limits_{S^2} \langle x,t|\Phi|x,t\rangle 
 = \int\limits_{\cM^{1,3}} \Omega\int\limits_{S^2} \langle x,t|\Phi|x,t\rangle 
 \  . 
\end{align}
As in section \ref{sec:trace}, it suffices to verify the latter for $SO(4)$ invariant functions $\Phi(x^0)$, and the semi-classical correspondence for slowly varying functions established above implies that $\Omega$ essentially coincides\footnote{The exact measure could be computed with some effort, which we leave to other work.} with the $SO(4,1)$ invariant measure \eqref{Omega-exact-E-M} on $H^4$.

The above (quasi-) coherent states can also be used to define a quantization map 
\begin{align}
  \cQ: \quad C^{\infty}(\C P^{1,2}) &\to \End(\cH_0)  \nn\\
   \phi &\mapsto \int\limits_{\C P^{1,2}} \tilde\Omega\, \phi(x,t) |x,t\rangle\langle x,t| 
    = \int\limits_{\cM^{1,3}} \Omega\int\limits_{S^2}  \phi(x,t) |x,t\rangle\langle x,t| 
   \label{Q-map-H4}
\end{align}
and a symbol map (de-quantization map) 
\begin{align}
  \End(\cH_0)
  &\mapsto C^{\infty}(\C P^{1,2})\nn\\
  \Phi &\mapsto \langle x,t|\Phi|x,t\rangle \ .
   \label{Q-map-H4-symbol}
\end{align}
These maps are manifestly compatible with $SO(4)$, and 
they are approximate inverse to each other in the semi-classical regime.
Equivariance under the non-compact transformations in $SO(4,1)$ is not manifest, but should hold to a good approximation at least in the semi-classical regime. 
Of course
these maps apply to functions on space-time $\cM^{1,3}$ by 
restricting to modes which are trivial\footnote{This is the subspace $\cC^0 \subset\End(\cH_0)$  characterized by the vanishing spin Casimir $\cS^2 = 0$ \cite{Sperling:2019xar}.} on $S^2$.


\subsection{Coherent states on $S^2_N$ from the oscillator construction}
\label{sec:fuzzy-S2-coh}

The above canonical coherent states can be illustrated for the fuzzy sphere $S^2_n$, which is constructed via 2 bosonic oscillators $a_1, a_2$ viewed as quantized complex functions on $\C^2_\theta$.
Consider a canonical coherent state on $\C^2_\theta$ located e.g. at $\vec \a = (0,\sqrt{n})$, given by
\begin{align}
    |\a\rangle = e^{-|\a|^2/2} e^{\a a^\dagger} |0\rangle \ .
\end{align}
The uncertainty in the space $\C^2$ is $\Delta a_i=1$, leading to an uncertainty of radius
$\Delta (a^*a) \approx a \Delta a \approx \sqrt{n}$.
The embedding function (i.e. the matrix background) for a fuzzy sphere $S^2_n$ with radius $n$ is given by angular momentum generators $L_i =  a^* \sigma_i a$.
Their expectation values defines a vector of length $\langle L_i\rangle \sim a^2 \sim n$, and
their uncertainty is of order $\Delta L_i \approx a \Delta a \sim \sqrt{n} = \frac{1}{\sqrt{n}}|L|$. 
The coherent states $|0\rangle_n$ on $S^2_n$ are given precisely by the projection of $|\a\rangle$ to $\hat N=n$, since
\begin{align}
    |\a\rangle = \sum_m c_m |0\rangle_m, 
\end{align}
where $c_m$ is approximately a Gaussian peaked around $n$ with uncertainty $\sqrt{n}$.
The uncertainty of $L_i$ is not affected by the projection, since the $L_i$ commute with $\hat N$.
Therefore the standard properties of coherent states on $S^2_n$ are inherited from the coherent states on $\C^2_\theta$.

\subsection{Quasi-coherent states: numerical approach}
\label{sec:quasicoh-num}

In this section, we show that the above coherent states can essentially be recovered using the more general construction of quasi-coherent states given in \cite{Steinacker:2020nva}.

Starting with the Hilbert space constructed in the Appendix \ref{sec:hilbertspace}, we construct quasi-coherent states as  ground states of the following displacement Hamiltonian \cite{Steinacker:2020nva}
\begin{equation}
     H_{( \vec{\bar{x}},\vec{\bar{t}} )}= \sum_{i=1}^3 (X^i - \bar x^i)^2+ (T^i - \bar t^i)^2+(X^4-\bar x^4)^2 \ .
\end{equation}
Up to $SO(4)$ rotations, we can restrict ourselves to localize the coherent states as in the analytical approach on\footnote{Note that once the point is fixed to lie on the "$z$-axis", $\bar x^4$ and $\bar t^\mu$ are completely fixed by $\bar x^\mu$, because of the constraints $\bar x^\mu \bar x_\mu=r^2=-\bar t^\mu\bar t_\mu $ and $\bar x^\mu \bar t_\mu=0$.} 
\begin{equation}
    \bar x^a=\begin{pmatrix}
       \bar x^0\\ 0\\ 0\\ \bar x^3\\ \bar x^4
    \end{pmatrix},\ \     \bar{t}^\mu= \frac{1}{r^2}\begin{pmatrix}
    \bar x^3  \\ 0\\ 0\\ \bar x^0
    \end{pmatrix}, \ \ \bar x^0\geq \bar x^3-r\gg r,\ \bar x^4\approx \sqrt{r^2+(\bar x_0)^2-(\bar x_3)^2} \ .
\end{equation}
With this ansatz, the problem is reduced to a subspace of $\mathcal{H}$ composed of $O(2)$ invariant states denoted by, $\ket{m,m}_n$, satisfying $J_3\ket{m,m}_n=0$, following the construction in Appendix \ref{sec:hilbertspace}. Such states localize (as confirmed numerically) on the highest weight states $\ket{n,n}_n$, exhibiting generically a Gaussian behavior. In fact, the Hamiltonian above
reduces 
in the specified regime to a Harmonic oscillator-like problem.\\

We find the ground states, i.e. coherent states, numerically to be
\begin{equation}
    \nket{\bar x,\bar t}=c\sum_n\, e^{-i\varphi n}\exp\left(-\frac{(n-\bar x^0)^2}{4\sigma^2}\right)\ket{n,n}_n,\ c=\left(2\pi \sigma\right)^{-\frac{1}{4}},\ \varphi=\arccos\left(\frac{\bar{x}_3}{r}\right)
\end{equation}
with $\sigma^2\sim  O(\bar x^0)$.\\

\begin{figure}[h!]
    \centering 
    \begin{overpic}[scale=.9]{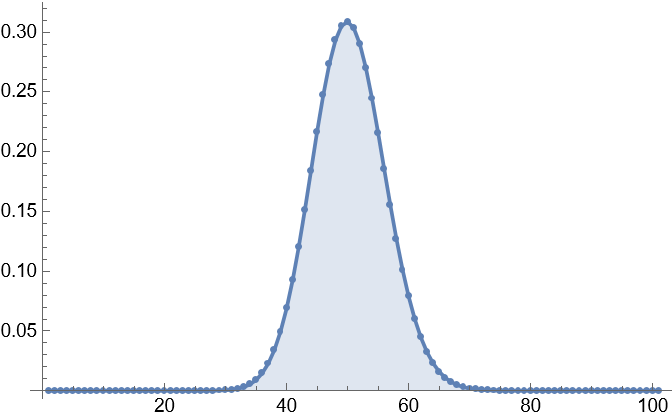}
\put(100,0){$n$}
\put(-2,65){$|c_n|^2$}
\put(65,50){$\nket{\bar x,\bar t}=\sum_nc_n\ket{n,n}_n$}
\end{overpic}
\end{figure}
\begin{figure}[h!]
  \centering 
\vspace{0.5cm}
    \begin{overpic}[scale=.9]{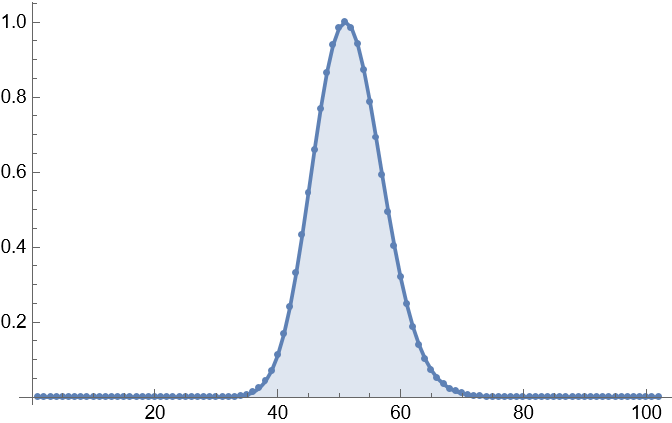}
\put(100,0){\footnotesize{$\frac{\bar x^0}{r}$}}
\put(-10,64){\footnotesize{$|\nbraket{\bar x^\prime,\bar t^\prime}{\bar x,\bar t}|^2$}}
\label{overlap}
\end{overpic}
\caption{Overlap of quasi-coherent states localized at different points on $\cM^{1,3}$. The relative uncertainty decreases with time.}
\end{figure}

For these states, the expectation value of $X^a$ and $T^\mu$ is found to be
\begin{align}
     \left\langle X^a\right\rangle
     &\approx \bar x^a  \qquad 
      \left\langle T^\mu\right\rangle\approx \bar t^\mu
\end{align}
consistent with \eqref{exp-values-coh-analytical}.
This means that the state is localized in $\cM^{1,3}$ and $H^4$ at 
\begin{align}
    x^a = \bar x^a,\ \ t^\mu =\bar t^\mu
\end{align}
respectively, with 
\begin{align}
0 > r^{-2} x_\mu x^\mu &= -r^{2}\bar x_0^2 + r^{-2} \bar x_3^2 \approx - r^2 t_\mu t^\mu \nn\\
x_a x^a &\approx r^2 \approx 0 \nn\\
x_\mu t^\mu &\approx 0
\label{semiclass-constraints}
\end{align}
in agreement with the constraints\footnote{Note that $x_a x^a \approx \pm r^2 \approx 0$ up to uncertainties, so that the sign is meaningless here.}\eqref{TT-XX-relation},  \eqref{TX-id} and  \eqref{radius-constraint}.
The variances are
\begin{equation}
    (\Delta X^\mu)^2 = O(r \bar x^0) \sim r^4 (\Delta T^\mu)^2
\end{equation}
consistent with  the uncertainly scale \eqref{LNC}
\begin{align}
    \Delta x^\mu \approx L_{\rm NC}, \qquad 
    \Delta t^\mu = r^{-2} L_{\rm NC} \ 
\end{align}
consistent with the analytic results of section \ref{sec:coh-states-osc}.
In particular, the {\em relative uncertainty}
\begin{align}
    \frac{\Delta\bar x}{\bar x^0} \sim \sqrt{\frac r{\bar x^0}} \sim e^{-\tau/2} \ \to 0
\end{align}
goes to zero at late times\footnote{the relation with $\tau$ holds near  the reference point \eqref{ref-point}.}.
These uncertainties are verified by computing the overlap between the coherent states centered at different points, as shown in figure \ref{overlap}.

Using the $SO(4)$ symmetry, these states can be moved to any location on the surface $\tilde S^3_R$ \eqref{Cauchy-surface-x0}.
All these results are consistent with the properties of the canonical states discussed in  section \ref{sec:coh-states-osc}, and we conclude that
\begin{align}
    \nket{x,t} \approx \ket{x,t} \ .
\end{align}
This justifies the semi-classical replacements
\begin{align}
 X^a \to x^a, \qquad T^\mu \to t^\mu 
\end{align}
satisfying the semi-classical constraints \eqref{constraints-approx}, with vanishing relative uncertainties 
\begin{align}
\frac{\Delta X^\mu}{\langle X^0\rangle} \sim \sqrt{\frac r{\bar x^0}} \ \ \to \ 0
\end{align}
in the late-time regime. Therefore minimal $\cM^{1,3}_0$ can be used as a semi-classical model for space-time, which is the main result of the present paper.

\section{String modes}
\label{sec:string-modes}

Coherent states are important not only to describe the semi-classical geometry, but also for evaluating loop computations for field theory on quantum spaces \cite{Steinacker:2022kji,Steinacker:2024unq}. For example, the one-loop effective action is given in terms of a trace over $\End(\cH)$. This can be computed using 
string modes, which are defined in terms of the above quasi-coherent states as follows 
\begin{align}
    \left|^{x,t}_{y,t'} \right) \equiv 
    \ket{x,t}| \bra{y,t'} \quad \in \End(\cH) \ .
\end{align}
By construction, all string modes are  covariant under $SO(4)$.
Then the trace over $\End(\cH)$
can be computed via an exact formula with the structure 
\begin{align}
    \Tr_{\End(\cH)} \cO = 
    \int 
    \left(^{x,t}_{y,t'} \right|
    \cO
    \left|^{x,t}_{y,t'} \right)
\end{align}
for $\cO$ some kinetic operator on  $\End(\cH)$, such as $f(\Box)$ or more complicated operators.
Here contributions with $x\approx y$ correspond to local contributions. Non-local contributions for $x\neq y$ arise 
e.g. due to UV/IR mixing, and are tamed in the maximally supersymmetric IKKT model; see \cite{Steinacker:2024unq} for a systematic discussion.

From the point of view of field theory on $\cM^{1,3}$, the presence of an extra label $t\in S^2$ is unexpected. This allows to capture the higher-spin modes arising from non-trivial harmonics on the internal $S^2$ fiber. It is possible to restrict to the sector $\cC^0$ \eqref{phi-hs-expansion} of "ordinary" functions on space-time  $\cM^{1,3}$ by averaging over $S^2$; for example,
\begin{align}
\label{position-states}
|x\rangle\langle x| := \int\limits_{S_t^2}
\ket{x,t}| \bra{x,t} 
\end{align}
could be considered as an optimally localized state on space-time $\cM^{1,3}$,
being constant along the fiber. 
Note however that this is a mixed state with rank larger than one, hence the notation \eqref{position-states} should be used with caution.
Similarly,
\begin{align}
[\Phi]_0(x) = c\int\limits_{S^2}\langle x,t|\Phi|x,t\rangle 
\end{align}
extracts the classical value of $\Phi\in\End(\cH)$ at $x\in\cM^{1,3}$ (for suitable $c$), cf. \eqref{kappa-def}.


\section{Conclusion and outlook}

This paper provides a self-contained discussion of minimal covariant quantum space-time. Our results  
 imply that minimal covariant quantum space-time $\cM^{1,3}_0$ fits into the framework of almost-local quantum geometries \cite{Steinacker:2020nva}, and can be interpreted as 
 $S^2$-bundle over space-time, or as
 quantized twistor space $\C P^{1,2}$. 
 The effective late-time $k=-1$ FLRW space-time geometry, with cosmic scale function $a(t)$, is the same as for the generic spaces $\cM^{1,3}_n$ for $n \gg 0$ \cite{Sperling:2019xar} ff, and is therefore not repeated here. 
  Since $\cM^{1,3}_0$ is shown to carry also a (locally finite) tower of $\hs$ modes,  the calculation of the one-loop effective action of the IKKT model given in \cite{Steinacker:2021yxt,Steinacker:2024huv,Manta:2024vol}  on $\cM^{1,3}_0 \times \cK_N$  (here  $\cK_N$ is some transversal compact fuzzy space) for $n\gg 0$ also applies to the minimal $n=0$ case, with minor adaptions.
   This one-loop action  is UV finite, and includes an  Einstein-Hilbert term. All degrees of freedom required for gravity
   do arise also for $n=0$, hence
 there is no compelling reason to consider non-minimal spaces with $n > 0$.

 Due to the large symmetry group and its simplicity, 
 this space is 
 one of the most promising candidates for realistic quantum space-time. The evolution of the cosmic scale $a(t)$ can be modified by including a time-dependent factor in the matrix background,  along the lines of \cite{Battista:2023glw}.

 The explicit characterization \eqref{XX-CR-explicit} in terms of generators and relations should provide a useful starting point for further work towards clarifying the  physics arising on this space-time.

\subsection*{Acknowledgement}

We would like to thank Tung Tran for collaboration in an early stage of this project. HS would also like to thank Pei-Ming Ho and Hikaru Kawai for useful discussions in a related collaboration.
This work is supported by the Austrian Science Fund (FWF) grant P36479.

\appendix

\section{Appendix}

\subsection{Proof of \eqref{cM-X-T-id-2}}
\label{sec:appendix-B}

We  determine the coefficients in \eqref{cM-X-T-id-1}:
\begin{align}
    \cM^{\mu\nu} = c(X^4)(T^\mu X^\nu - T^\nu X^\mu) 
     + d(X^4) \varepsilon^{\mu\nu\rho\sigma} T_\rho X_\sigma \ .
\end{align}
The quadratic Casimir \eqref{SO31-casimir-1} then gives 
\begin{align}
  -1-r^{-2} X_4^2 &= \frac 12 \cM^{\mu\nu} \cM_{\mu\nu} \nn\\
  &= \frac 12 c(X^4)(T^\mu X^\nu - T^\nu X^\mu) \cM_{\mu\nu}
     + \frac 12 d(X^4) \varepsilon^{\mu\nu\rho\sigma} T_\rho X_\sigma
 \cM_{\mu\nu}  \nn\\
 &= c(X^4) T^\mu X^\nu \cM_{\mu\nu}
  = c(X^4) T^\mu (2i X_\mu -  r X_4 T_\mu) \nn\\
  &= -c(X^4) (4 r^{-1} X^4 + r T^\mu  X_4 T_\mu)\nn\\
  &= -c(X^4) (4 r^{-1} X^4 + r T^\mu  (i r^{-1} X^\mu + T_\mu X_4))\nn\\
  &= -c(X^4) (4 r^{-1} X^4 + i T^\mu X^\mu + r T^\mu T_\mu X_4)\nn\\
  &= -r^{-1} c(X^4) (X^4 + r^{-2} X_4^3)  \nn\\
  c(X^4) &= r X_4^{-1}
\end{align}
using 
\begin{align}
0 &= \cM^{\mu b} X_b + X_b\cM^{\mu b} 
 = 2X_b\cM^{\mu b} - 4i X^\mu  \nn\\
 &= 2X_\nu\cM^{\mu \nu}  + 2 r X_4   T^{\mu} - 4i X^\mu \nn\\
 0 &= T_\mu X^\mu + X^\mu T_\mu = 2 T_\mu X^\mu -4 i r^{-1} X^4
\end{align} 
with \eqref{MX-id} and $\cM^{ab} X_b = -4i X^a + X_b \cM^{ab}$. 
Furthermore, \eqref{SO31-casimir-2} says
\begin{align}
   0 &= \varepsilon_{\mu\nu\rho\sigma} \cM^{\mu\nu} \cM^{\rho\sigma} 
   = \varepsilon_{\mu\nu\rho\sigma}( c(X^4)(T^\mu X^\nu - T^\nu X^\mu) 
     + d(X^4) \varepsilon^{\mu\nu\rho\sigma} T_\rho X_\sigma)
   \cM^{\rho\sigma} \nn\\
   &=d(X^4) \varepsilon_{\mu\nu\rho\sigma} \varepsilon^{\mu\nu\rho'\sigma'} T_{\rho'} X_{\sigma'}
   \cM^{\rho\sigma}  \nn\\
   &= d(X^4)  T_\rho X_\sigma \cM^{\rho\sigma} 
    =  -2r^{-1} d(X^4) (X^4 + r^{-2} X_4^3) 
\end{align}
hence $d(X^4) = 0$.

\subsection{Group-theoretic coherent states:  
$SO(4,1)$ versus $SO(4,2)$}

Since $\cH = \cH_0$ is a lowest weight representation of $SO(4,2)$, 
there are natural ("Perelomov") coherent states defined 
in terms of the $SO(4,2)$ or $SO(4,1)$ orbit of the
lowest weight state $|\Lambda\rangle \in \cH$.
However, these are {\em not} satisfactory in the present context, for the following reason:

\paragraph{$SO(4,1)$ point of view.}

Since $\cH_0$ is irreducible under $SO(4,1)$, it suffices to consider only the  $SO(4,1)$ orbit  of $|\Lambda\rangle$
to obtain an over-complete basis\footnote{Note that $\ket\L$ cannot be considered as lowest weight state of $SO(4,1)$, so that no statements about holomorphicity apply to $H^4$.} of $\cH_0$. 
This $SO(4,1)$ orbit  is a 4-dimensional manifold (up to phase),
generated by the 4 generators $\cM^{0a}$ for $a=1,...,4$
\begin{align}
  \cM :=  SO(4,1)|\Lambda\rangle/_{U(1)} \cong H^4
\end{align}
since $|\Lambda\rangle$ is invariant under $SO(4)$. 
This defines an "abstract quantum space" \cite{Steinacker:2020nva}, which is isomorphic to a 4-dimensional hyperboloid $H^4$.
More explicitly,
let $g_x \in SO(4,1)$ be some group element that rotates the "north pole" 
\begin{align}
    \tn=r(1,0,0,0,0) \in \R^{1,4}
\end{align}
to a generic point $x = g_x\cdot \tn  \in H^4$. Let $U_x := \Pi(g_x)$ be the corresponding unitary operator on $\cH$. Then 
\begin{align}
\label{coherent-covariance}
 |x\rangle \coloneqq U_x |\Lambda\rangle  \,, \qquad x = g_x \cdot \tn
  \; .
\end{align}
We shall denote these as "group-theoretic" coherent states.
Accordingly, these states fail to  resolve the internal $S^2$; this failure is remedied by the quasi-coherent states discussed in section \ref{sec:coh-states-osc}.
For the doubleton representations $\cH_n$ with  $n\gg 0$, 
the analogous group-theoretic coherent  states do resolve the internal $S^2$ and hence the bundle geometry up to spin $n$  \cite{Sperling:2018xrm}.

\paragraph{$SO(4,2)$ point of view.}

Now consider the orbit of $SO(4,2)$ on $\ket{\L}$
\begin{align}
\cM_\C = SO(4,2)|\Lambda\rangle/_{U(1)} \ .
\end{align}
This is an 8-dimensional manifold (up to phase),
generated by the 8 generators 
$\cM^{0a}$ and $\cM^{a5}$ for $a=1,...,4$.
Since $Z_a^-\ket{\L} = (\cM_{a5} - i \cM_{a0})\ket{\L} = 0$, some of 
these are related by $i$
\begin{align}
   r^{-1} X^a\ket{\L} = i \cM^{0a} \ket{\L}
\end{align}
i.e. $X^j\ket{\L} = i V^j \ket{\L}$;
however they are independent from the real point of view.
Hence the $SO(4,2)$ orbit of coherent states is 8-dimensional, and it is {\bf not}  isomorphic\footnote{This was not  appreciated in the first version of \cite{Sperling:2018xrm} ff., so that the statements about truncated $\hs$ towers apply only to a "holomorphic" sub-sector of $\End(\cH)$.} to $\C P^{1,2}$. 



\subsection{Expectation values associated with group-theoretic coherent states}\label{sec:cohe-uncertain}

\paragraph{Expectation values.} 

For the lowest weight state located at the north pole $\tn$, the following expectation values hold:
\begin{subequations}
    \begin{align}
 \langle \L|X^a|\L\rangle &= r \delta^{a,0}\, & a=0,\ldots,4\,,  \nn\\
\langle \L|T^\mu|\L\rangle &= 0\, &\mu=0,1,2,3\,   \nn\\
\langle \L|\cM^{\mu\nu}|\L\rangle &= 0  \, , &\mu=0,1,2,3\,.
\label{eq:MM-so(3,1)-EV}
\end{align}
\end{subequations}
The identity $\langle\L|T^0|\L\rangle = 0$ follows from $T^0 \propto [X^0,X^4]$ together with $X^0|\Lambda\rangle = r |\Lambda\rangle$. 
Furthermore observing  $[\cM^{ij},X^0] = 0 = [T^{i},X^0]$ we obtain
    $\langle \L|\cM^{ij}|\L\rangle = 0 = \langle \L|T^{i}|\L\rangle$ because $|\L\rangle$ is the unique ground state of $X^0$. Finally, $\cM^{0i} \sim [X^0,X^i]$ together with 
$X^0 |\L| = r|\L\rangle$ implies $\langle \L|\cM^{i0}|\L\rangle = 0$.

Using the covariance property \eqref{coherent-covariance} under $SO(4,1)$, this gives immediately the expectation values of some simple operators associated with the generic coherent states $|x\rangle$:
\begin{subequations}
    \begin{align}
 \langle x|X^a|x\rangle &= x^a\,, \qquad &a&=0,\ldots,4\,,  \nn\\
  \langle x|T^{\mu}|x\rangle&=0\,,\qquad &\mu&=0,1,2,3\, \\
  \langle x|\cM^{ab}|x\rangle &= 0 ,\qquad  &a,b&=0,\ldots,4\,. \label{Mab-expect}
\end{align}
\end{subequations}
Now consider quadratic operators. The radius constraint \eqref{radius-constraint} together with $SO(4)$ invariance leads to
\begin{equation}
\label{quadratic-expect-basic-X}
 \bra{\Lambda}(X^0)^2\ket{\Lambda}=r^2,\qquad   \bra{\Lambda}(X^a)^2\ket{\Lambda}=\frac{r^2}{2}\, \qquad a=1,...,4 \ .
\end{equation}
Similarly, we can compute the expectation values of the $(T^\mu)^2$. We first observe  
$T^i|\L\rangle = r^{-1}\cM^{i4}|\L\rangle = 0$, since $|\L\rangle$ is invariant under $SO(4)$.
This implies that $ \bra{\Lambda}T_i^2\ket{\Lambda}=0$. Then 
the constraint \eqref{Tsquare-id} together with 
\eqref{quadratic-expect-basic-X} gives 
$\bra{\Lambda}(T^0)^2\ket{\Lambda}= \frac{1}{r^2} -  \bra{\Lambda}(X^4)^2\ket{\Lambda} = \frac{1}{2r^2}$.
Thus
\begin{equation}
\label{quadratic-expect-basic-T}
 \bra{\Lambda}(T^0)^2\ket{\Lambda}=\frac 1{2r^2},\qquad   \bra{\Lambda}(T^i)^2\ket{\Lambda}=0\, \qquad i=1,2,3 \ .
\end{equation}
Now we want to obtain these expectation values at generic points $x=g_x\cdot \tn \in H^4$.
Due to the  $SO(1,3)$ invariance, it suffices to consider the reference point 
\begin{align}
\label{ref-point-full}
    \xi = (x^0,0,0,0,x^4) =  r(\cosh(\tau),0,0,0,\sinh(\tau))\quad \in H^4 \ 
\end{align}
which is obtained by acting with $e^{-i \tau \Sigma^{04}}$ on $\tn$.
Hence the corresponding coherent state is
\begin{align}
\label{U-tau}
   |x\rangle &= U_\tau |\L\rangle, \qquad 
    U_\tau =  e^{-i \tau\cM^{04}} \ .
\end{align}
To proceed, we need the following transformation formulas
for boosts in the $(X^0,X^4)$ plane 
\begin{align}
U_{\tau}^{-1}X^0 U_\tau
&= X^0\cosh\tau + X^4\sinh\tau , \qquad
    U_{\tau}^{-1} X^i U_\tau = X^i , \qquad i=1,2,3  \nn\\ 
U_{\tau}^{-1}X^4 U_\tau
&= X^0\sinh\tau + X^4\cosh\tau  \\
    U_{\tau}^{-1} T^0 U_\tau &= T^0 , \qquad 
    U_{\tau}^{-1} T^i U_\tau =\cosh(\tau) T^i - \sinh(\tau) r^{-1} \cM^{0i} \ .
    \label{Utau-trafos}
\end{align}
Together with the algebraic constraints,
this gives 
\begin{align}
\label{quadratic-expectations-basic-x}
   \bra{x} (X^0)^2 \ket{x}
 &= r^2 \big(1 + \frac{3}{2} \sinh^2\tau\big) , 
  &\bra{x} (X^i)^2 \ket{x}
= \frac{r^2}{2}  \nn\\
   \bra{x} (X^4)^2 \ket{x}
 &= \frac {r^2}2 \big(1+3\sinh^2\tau\big) &  \nn\\
 \bra{x} (T^i)^2 \ket{x}
   &=  \frac{1}{2r^2} \sinh^2\tau, 
  &\quad \bra{x} (T^0)^2 \ket{x} 
= \frac{1}{2r^2} \ .
\end{align}

\paragraph{Uncertainties of $X^a$ and $T^\mu$.}
Using the above results, we obtain the uncertainties w.r.t. $|x\rangle$ at the reference point
\begin{align}
    \Delta X^0 &=   \frac{r}{\sqrt{2}} |\sinh\tau|, \qquad
  \Delta X^i = \frac{r}{\sqrt{2}}  \qquad
    \Delta X^4 = \frac{r}{\sqrt{2}} \cosh\tau     \nn\\
     \Delta T^0 &=\frac{1}{\sqrt{2}r} , \qquad\qquad
     \Delta T^i =\frac{1}{\sqrt{2}r} |\sinh\tau| \ .
\end{align}
We observe that the uncertainty of $X^i$ is very small (at late times $\tau)$, but the uncertainty of $X^0$ and $X^4$ is very large,  comparable with their expectation values. This means that the group-theoretic 
coherent states $|x\rangle$ are very badly balanced, and do not resolve the geometry.

\subsection{On the structure of $\mso(4,2)$}
\label{sec:mso42-structure}

The maximal compact subgroup of $SO(4,2)$ is given by  $SO(4)$ with generators $\cM^{ab}$, $a,b=1,...,4$ and the $O(2)_H$ generated by 
\begin{align}
    H = r^{-1} X^0 =\cM^{05} \ .
\end{align}
We define the $SO(4)$ vector operators as follows:
\begin{align}
 X^a := \cM^{a 5}, \qquad V^a := \cM^{a 0}, \qquad a,b=1,2,3,4
\end{align}
(setting $r=1$ in this section),
which mix under $SO(2)_H$.
They satisfy
\begin{align}
 [X^a,X^b] &= i \cM^{ab},  \nn\\
 [V^a,V^b] &= i \cM^{ab},  \nn\\
 [V^a,X^b] &= i \delta^{ab} H \ .
 \label{X-P-CR}
\end{align}

\paragraph{Root generators.}

\vspace{0.2cm}

\begin{figure}
\begin{center}
 \includegraphics[width=0.5\textwidth]{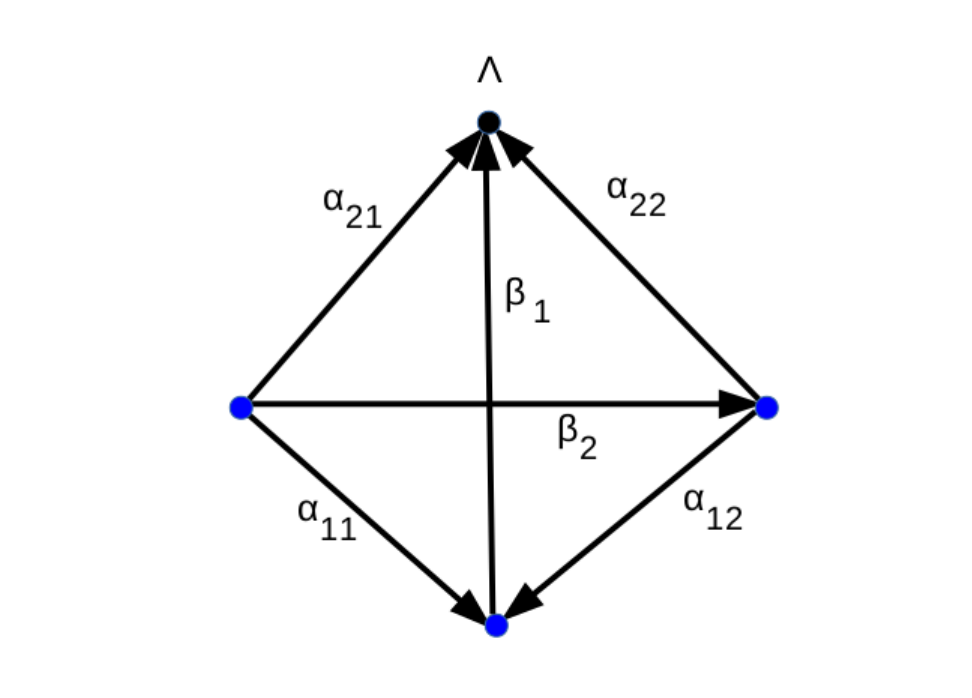}
 \end{center}
 \caption{Tetrahedral structure of $\mso(6) = \msu(4)$ roots in weight space $\R^3$. $X^0 = H$ is the Cartan generator perpendicular to the compact  roots $\b_i$, cf. fig. \ref{fig:weights-minirep}.}
 \label{fig:SUU4-structure}
\end{figure}

It is  instructive to understand the structure in terms of roots and weights.
The Lie algebra $\mso(4,2) \cong \msu(2,2)$ has 6 roots and 3 Cartan generators, which we can choose to be 
$\{H,L_3,T_3\}$ where $L_3 := \cM^{12}$
and $T_3 =\cM^{34}$.
We organize the roots as sketched in figure \ref{fig:SUU4-structure}:
We single out the compact $SO(4) = SU(2)_{L} \times SU(2)_{R}$ subgroup 
corresponding to the two orthogonal roots $\b_1$ and $\b_2$. These two subgroups commute, hence
\begin{align}
 [X_{\b_1}^\pm, X_{\b_2}^\pm] = 0 \ .
\end{align}
Then the remaining generators $X_{\a_{ij}}^\pm$ for $i,j=1,2$ transform as complex vectors 
$(4)_\C = (2)_{\b_1} \otimes (2)_{\b_2}$ under  $SO(4)$.
The commutation relations can be written in terms of
\begin{align}
 Z_a^\pm = (\sigma_a)^{ij} \, X_{\a_{ij}}^\pm 
\end{align}
which transform as $(4)$ under $SO(4)$.
We recover \eqref{X-P-CR} upon
separating these into hermitian and anti-hermitian parts
\begin{align}
 Z_a^\pm = X_a \pm i V_a \ .
\end{align}

\subsection{Hilbert Space and Oscillator Construction}\label{sec:hilbertspace}
  Following \cite{Sperling:2018xrm}, the Hilbert space for the representation $\mathcal{H}_0$ can be built in an oscillator construction $SU(2)\times SU(2)$ construction, as illustrated in section 1.
  Each $SU(2)$ factor has two annihilation and creation operators $a_i,a_i^\dagger,b_i,b_i^\dagger$, $i=1,2$. A basis of $\mathcal{H}_0$  is constructed by acting on the unique  vacuum $\ket{\Lambda}$ (corresponding to the point in the hyperboloid $(r,0,0,0,0)$) annihilated by $a_i,b_i$ with the creation operators as follows
    \begin{equation}
        \ket{m_{a_1},m_{b_1}}_n:=\frac{(a^\dagger_1)^{m_{a_1}}(a^\dagger_2)^{n-m_{a_1}}(b^\dagger_1)^{m_{b_1}}(b^\dagger_2)^{n-m_{b_1}}}{\sqrt{m_{a_1}!(n-m_{a_1})!m_{b_1}!(n-m_{b_1})!}}\ket{\Lambda},\ \ \ m_{a_1},m_{b_1}\in\{0,1,\ldots,n\}
    \end{equation}
   for $n\geq 0$, with $\ket{\Lambda}\equiv \ket{0,0}_0$. These states form the subspace of Fock space restricted to  $N_b-N_a=0$.
The number $n$ then corresponds to the $X^0$ eigenvalues, by $X^0\ket{m_a,m_b}_n=r(n+1)\ket{m_a,m_b}_n$.

In terms of these oscillator ladder operators, the hermitian operators $\mathcal{M}^{ab}$ are given by
\begin{align}
\label{osc-ops-explicit}
    \mathcal{M}_{01}&=\frac{1}{2i}(-a_1b_2-a_2b_1+b_1^\dagger a_2^\dagger+b_2^\dagger a_1^\dagger)\\
    \mathcal{M}_{02}&=\frac{1}{2}(a_1b_2-a_2b_1-b_1^\dagger a_2^\dagger+b_2^\dagger a_1^\dagger)\\
    \mathcal{M}_{03}&=\frac{1}{2i}(-a_1b_1+a_2b_2+b_1^\dagger a_1^\dagger-b_2^\dagger a_2^\dagger)\\
    \mathcal{M}_{12}&=-\frac{1}{2}(-a_1a_1^\dagger+a_2a_2^\dagger+b_1^\dagger b_1-b_2^\dagger b_2)\\
    \mathcal{M}_{23}&=-\frac{1}{2}(-a_1a_2^\dagger-a_2a_1^\dagger+b_1^\dagger b_2+b_2^\dagger b_1)\\
    \mathcal{M}_{31}&=-\frac{i}{2}(a_1a_2^\dagger-a_2a_1^\dagger-b_1^\dagger b_2+b_2^\dagger b_1)\\
    \mathcal{M}_{04}&=\frac{1}{2}(a_1b_1+a_2b_2+b_1^\dagger a_1^\dagger+b_2^\dagger a_2^\dagger)\\
    \mathcal{M}_{14}&=\frac{1}{2}(a_1a_2^\dagger+a_2a_1^\dagger+b_1^\dagger b_2+b_2^\dagger b_1)\\
    \mathcal{M}_{24}&=\frac{i}{2}(-a_1a_2^\dagger+a_2a_1^\dagger-b_1^\dagger b_2+b_2^\dagger b_1)\\
    \mathcal{M}_{34}&=\frac{1}{2}(a_1a_1^\dagger-a_2a_2^\dagger+b_1^\dagger b_1-b_2^\dagger b_2)\\
    \mathcal{M}_{05}&=\frac{1}{2}(a_1a_1^\dagger+a_2a_2^\dagger+b_1^\dagger b_1+b_2^\dagger b_2)\\
    \mathcal{M}_{15}&=\frac{1}{2}(a_1b_2+a_2b_1+b_1^\dagger a_2^\dagger+b_2^\dagger a_1^\dagger)\\
    \mathcal{M}_{25}&=\frac{i}{2}(-a_1b_2+a_2b_1-b_1^\dagger a_2^\dagger+b_2^\dagger a_1^\dagger)\\
    \mathcal{M}_{35}&=\frac{1}{2}(a_1b_1-a_2b_2+b_1^\dagger a_1^\dagger-b_2^\dagger a_2^\dagger)\\
    \mathcal{M}_{45}&=\frac{i}{2}(-a_1b_1-a_2b_2+b_1^\dagger a_1^\dagger+b_2^\dagger a_2^\dagger) \ .
\end{align}
From these one can easily read off the $X^a=r\mathcal{M}^{a5}$, $T^a=\frac{1}{r}\mathcal{M}^{a4}$, the boosts $\mathcal{M}^{0a}$, and the rotations $J_i=\frac{1}{2}\varepsilon_{ijk}\mathcal{M}^{jk}$.

\bibliographystyle{JHEP}
\bibliography{twistor}

\end{document}